\documentclass[intlimits,twoside,a4paper]{article}

\usepackage{graphicx}			
\usepackage{amsmath}			
\usepackage{amsfonts}
\usepackage{amssymb}
\usepackage{amsthm}
\usepackage[T2A]{fontenc}
\usepackage[cp1251]{inputenc}

\usepackage[eqsecnum]{cmpj2}

\issue{2016}{19}{3}{33003}
\doinumber{10.5488/CMP.19.33003}

\title[Bethe approach study of the mixed spin-1/2 and spin-5/2 Ising system]
{ Bethe approach study of the mixed spin-1/2 and spin-5/2 Ising system in the presence of an applied
magnetic field}

\author[M. Karimou \textsl{et al.}]{M.~Karimou\refaddr{label1}, R.A.~Yessoufou\refaddr{label1,label2}\thanks{E-mail: yesradca@yahoo.fr, olorire2012@gmail.com}\,, T.D.~Oke\refaddr{label1,label2}, A.~Kpadonou\refaddr{label1,label3}, F.~Hontinfinde\refaddr{label1,label2}}
\addresses{
\addr{label1} Institute of Mathematics and Physical Sciences (IMSP), Republic of Benin
\addr{label2} Department of Physics, University of Abomey-Calavi, Republic of Benin
\addr{label3} University of Natitingou, \'{E}cole Normale Sup\'{e}rieure, Republic of Benin
}

\authorcopyright{M.~Karimou, R.A.~Yessoufou, T.D.~Oke, A.~Kpadonou, F.~Hontinfinde, 2016}
\date{Received January 18, 2016, in final form May 13, 2016}

\begin{document}
\maketitle

\begin{abstract}

The mixed spin-1/2 and
spin-5/2 Ising model is investigated on the Bethe
lattice in the presence of a magnetic field $h$ via the recursion relations method. A ground-state phase
diagram is constructed which may be needful to explore important regions of the temperature phase diagrams of a model.
The order-parameters, the corresponding response functions and internal energy are thoroughly
investigated in order to typify the nature of the
phase transition and to get the corresponding temperatures. So, in the absence of the magnetic field, the temperature
phase diagrams are displayed in the case of an equal crystal-field on the
$(k_{\textrm{B}}T/|J|, \,D/|J|)$ plane when $q=3,4$, $5$ and $6$. The model only exhibits the second-order
phase transition for appropriate values of physical parameters of a model.


\pacs 05.50.+q, 05.70.Ce, 64.60.Cn, 75.10.Hk, 75.30.Gw
\end{abstract}

\section{Introduction}

Over the last five decades, the Ising model has been one of the most largely used models to
describe critical behaviors of several systems in nature. Lately, numerous extensions have been made in
the \linebreak spin-$\frac{1}{2}$ Ising model to describe a wide variety of systems. For example, the models
consisting of mixed spins with different magnitudes are interesting extensions, forming
the so-called mixed-spin Ising class \cite{ra1,ra2}. Beyond that, magnetic materials have several important technological
applications: they find wide use in information storage devices, microwaves communication systems,
electric power transformers and dynamo, and high-fidelity speakers \cite{ra3,ra4,ra5,ra6}. Thus, in response to an increasing demand
placed on the performance of magnetic solids, there has been  a surge of interest in
molecular-based magnetic materials \cite{ra7,ra8,ra9,ra10}.
Indeed, the discovery of these materials \cite{ra11} has been one of the
advances in modern magnetism. Many magnetic materials have two types of magnetic atoms regularly
alternating which exhibit ferrimagnetism. In this context, a good description of their physical
properties is given by means of mixed-spin configurations. The interest in studying
magnetic properties of these materials is due to their reduced
translational symmetry rather than to their single-spin counterparts, since they consist of two
interpenetrating sublattices. Thus, ferrimagnetic materials are of great interest due to their
possible technological applications and from a fundamental point of view.
These materials are modelled using mixed-spin Ising models that can be built up by infinite combinations
of different spins.

In literature there exist many studies on mixed-spin Ising systems which intend to clarify the magnetic
properties of magnetic systems. In this regard, there has been great interest in the study of magnetic
properties of the systems formed by two sublattices with different spins and crystal-field interactions
\cite{ra12}. Theoretically, such systems have been widely
analysed using several  numerical approaches, e.g., effective-field theory \cite{ra13,ra14,ra15,ra16,ra17,ra18}, mean-field
approximation \cite{ra19,ra20,ra21,ra22,ra23}, renormalization-group technique \cite{ra24,ra25}, numerical
simulations based on Monte-Carlo \cite{ra26,ra27,ra28,ra29,ra30} and exact recursion equations
\cite{ra31,ra32,ra33,ra34,ra35,ra36,ra37,ra38}. A somewhat newer interest is to extend such
investigations into a more general mixed-spin Ising model with one constituent spin-$\frac{1}{2}$ and the
other constituent spin-$\frac{5}{2}$. To this end, Deriven et al. \cite{ra18} used an effective-field
theory with correlation to study the same model and got interesting results. Recently, Guo et
al. \cite{ra17} studied the thermal entanglement of the same model by means of the concept of negativity
and also got interesting results concerning the effects of the magnetic field on the entanglement.

In the present paper, we use the exact recursion equations technique to examine the magnetic properties
of the mixed spin-$\frac{1}{2}$ and spin-$\frac{5}{2}$ Ising model
with equal crystal-field on the Bethe lattice in the presence of a longitudinal magnetic field.
The aim of this work is to investigate the effect of the crystal-field and the
magnetic field on the physical magnetic properties of the model.

The remainder of this paper is arranged as follows. In section~\ref{sec2}, the description of the model on the
Bethe lattice is clarified. Furthermore, the order-parameters, the corresponding response functions and the internal
energy are expressed in terms of
recursion relations. In the next section, some definitions of the critical temperature of the model
are explained. In section~\ref{sec4}, we present some illustrations and discuss in detail the numerical results. We finally conclude in the
last section .
\section{Description of the model on the Bethe lattice}\label{sec2}

\begin{figure}[!b]
\begin{center}
 \includegraphics[width=0.5\textwidth]{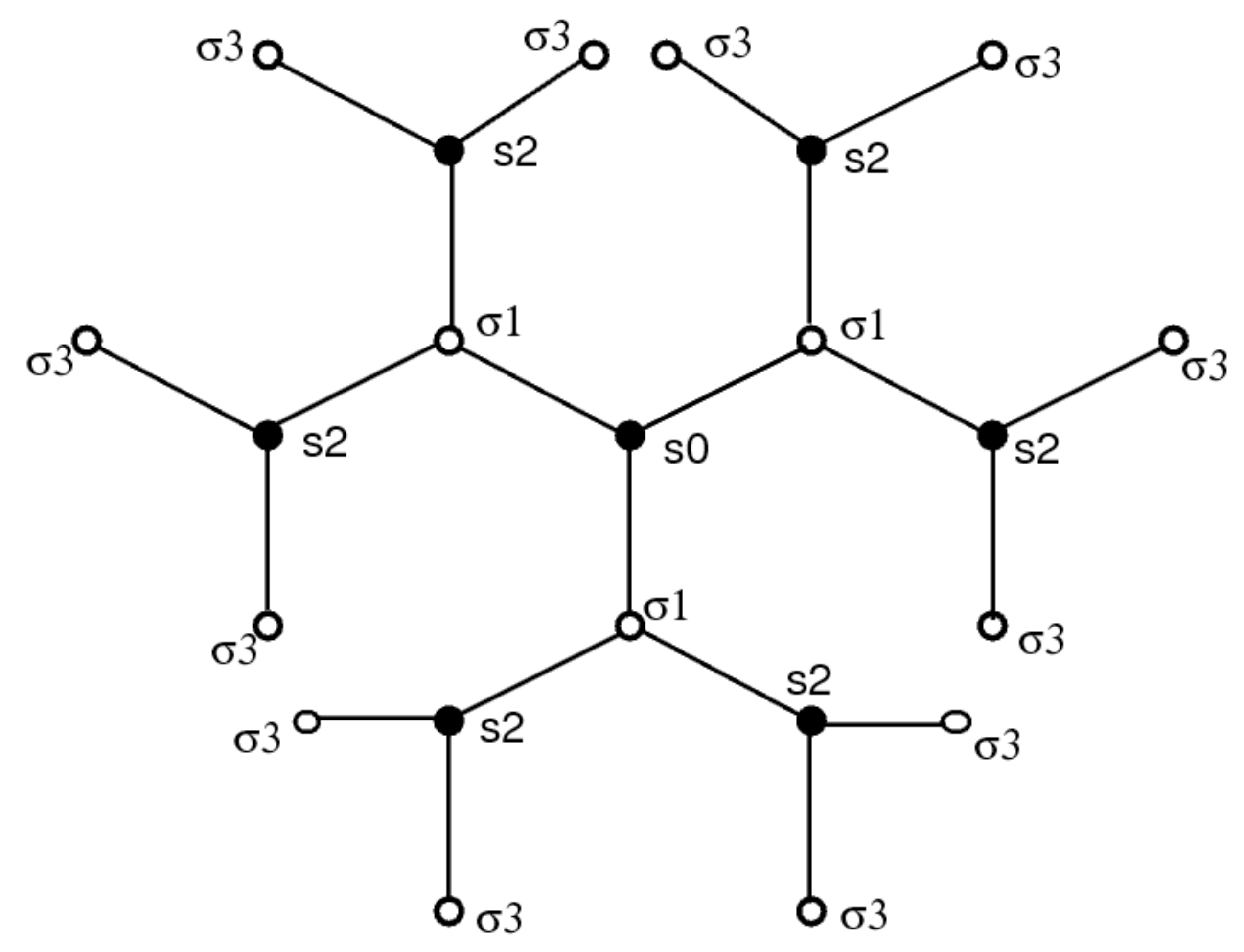} 
\end{center}
\caption{
The mixed spin Ising model consisting of two different magnetic atoms with spins values $s_i=\frac{1}{2}$
and $\sigma_j=\frac{5}{2}$, respectively, defined on the Bethe lattice with coordination number $q=3$.}
\label{fig1}
 \end{figure}

The mixed spins system on the Bethe lattice is shown in figure~\ref{fig1}.
We consider the mixed spin-$\frac{1}{2}$ and spin-$\frac{5}{2}$ system consisting of two sublattices A and
B. The sites of sublattice A are occupied by atoms of spins $S_i$, where $S_i=\pm{\frac{1}{2}}$. Those of the
sublattice B are occupied by atoms of spins $\sigma_j$, where
$\sigma_j=\pm{\frac{5}{2}}, \pm{\frac{3}{2}},\pm{\frac{1}{2}}$. In our case, the Bethe
lattice is arranged in such a way that the central spin is spin-$\frac{1}{2}$ and the next generation spin is
spin-$\frac{5}{2}$ and so on to infinity. Thus, the Ising  Hamiltonian of such a model on the Bethe
lattice may be written as:
 \begin{eqnarray}
 H =-J \sum_{\langle i,j\rangle}{\sigma_{j}}S_{i} -D \sum_{j}{\sigma_{j}^{2}}-h\left( \sum_{i}{S_{i}+ \sum_{j}\sigma_{j}}\right);
\end{eqnarray}
 $J<0$ is the bilinear exchange coupling interaction strength. $D$ and $h$ are, respectively, the single-ion
 anisotropy or the crystal-field and the longitudinal magnetic field acting on the
spins of the model.

In order to formulate the problem on the Bethe lattice, the partition function is the main ingredient which
 is given as:
\begin{eqnarray}
Z= \sum{\exp{\left\{\beta \left[ J \sum_{\langle i,j\rangle}{\sigma_{j}S_{i}} +D  \sum_{j}{\sigma_{j}^{2}} + h\left(\sum_{i}{S_{i}} + \sum_{j}{\sigma{_j}}\right) \right]\right\}}}.
\end{eqnarray}
If the Bethe lattice is cut at the central spin $S_0\,$, it splits into $q$  disconnected pieces. Thus,
the partition function on the Bethe lattice can be written as:
\begin{eqnarray}
 Z=\sum_{S_0}{\exp{\left[\beta ( h S_{0})\right]} g_{n}^{q}(S_0)},
\end{eqnarray}
where $S_{0}$ is the central spin value of the lattice, $ g_{n}(S_{0})$ is the partition function of an individual branch and the suffix $n$
represents the fact that the sub-tree has $n$ shells, i.e., $n$ steps from the root to  the boundary sites.
If we continue to cut the Bethe lattice on the sites $\sigma_1$ and $S_2$ which are respectively
the nearest and the next-nearest of the central spin $S_0\,$, we can obtain the recurrence relations
for $ g_{n}(S_{0})$ and $g_{n-1}(\sigma_{1}) $ as:
\begin{equation}
 g_{n}(S_{0})= \sum_{\{\sigma_{1}\}}\exp\left[\beta\left(JS_{0}\sigma_{1}+D\sigma_{1}^{2}+ h \sigma_1\right)\right]\left[g_{n-1}(\sigma_{1})\right]^{q-1},
\end{equation}
\begin{equation}
 g_{n-1}(\sigma_{1})= \sum_{\{S_{2}\}}\exp\left[\beta\left(JS_{2}\sigma_{1}+h S_{2}\right)\right]\left[g_{n-2}(S_{2})\right]^{q-1}.
\end{equation}

Now, we explicitly calculate some $g_{n}(S_{0})$ and $g_{n-1}(\sigma_{1})$ as follows:
 \begin{align}
    g_{n}\left(\pm\frac{1}{2}\right)&= \sum_{\{\sigma_{1}\}}\exp\left[\beta\left(\pm \frac{J}{2} \sigma_{1}+D\sigma_{1}^{2}+h \sigma_1\right)\right]\left[g_{n-1}(\sigma_{1})\right]^{q-1}\nonumber\\
  &= \exp{\left[\beta\left(\pm\frac{5J}{4}+\frac{25}{4}D +\frac{5}{2}h\right)\right]}\left[g_{n-1}\left(\frac{5}{2}\right)\right]^{q-1}
   + \exp{\left[\beta\left(\mp\frac{5J}{4} +\frac{25}{4}D -\frac{5}{2}h \right)\right]}\left[g_{n-1}\left(-\frac{5}{2}\right)\right]^{q-1}\nonumber\\
  &\quad + \exp{\left[\beta\left(\pm\frac{3J}{4} +\frac{9}{4}D +\frac{3}{2}h\right)\right]}\left[g_{n-1}\left(\frac{3}{2}\right)\right]^{q-1}
   +  \exp{\left[\beta\left(\mp\frac{3J}{4} +\frac{9}{4}D -\frac{3}{2}h\right)\right]}\left[g_{n-1}\left(-\frac{3}{2}\right)\right]^{q-1} \nonumber \\
  &\quad +  \exp{\left[\beta\left(\pm\frac{J}{4}+\frac{1}{4}D +\frac{1}{2}h\right)\right]}\left[g_{n-1}\left(\frac{1}{2}\right)\right]^{q-1}
  + \exp{\left[\beta\left(\mp\frac{J}{4} +\frac{1}{4}D -\frac{1}{2}h\right)\right]}\left[g_{n-1}\left(-\frac{1}{2}\right)\right]^{q-1},
  \end{align}
  \begin{align}
    g_{n-1}\left(\pm\frac{5}{2}\right)&= \sum_{\{S_{2}\}}\exp\left[\beta\left(\pm\frac{ 5J}{2}{S_{2}}+h S_{2}\right)\right]\left[g_{n-2}(S_{2})\right]^{q-1}\nonumber\\
  &= \exp{\left[\beta\left(\pm\frac{5J}{4}+\frac{h}{2}\right)\right]}\left[g_{n-2}\left(\frac{1}{2}\right)\right]^{q-1}
 + \exp{\left[\beta\left(\mp\frac{5J}{4}-\frac{h}{2}\right)\right]}\left[g_{n-2}\left(-\frac{1}{2}\right)\right]^{q-1},
 \end{align}
  \begin{align}
    g_{n-1}\left(\pm\frac{3}{2}\right)&= \sum_{\{S_{2}\}}\exp\left[\beta\left(\pm\frac{ 3J}{2}{S_{2}}+h S_{2}\right)\right]\left[g_{n-2}(S_{2})\right]^{q-1}\nonumber\\
  &= \exp{\left[\beta\left(\pm\frac{3J}{4}+\frac{h}{2}\right)\right]}\left[g_{n-2}\left(\frac{1}{2}\right)\right]^{q-1}
 + \exp{\left[\beta\left(\mp\frac{3J}{4}-\frac{h}{2}\right)\right]}\left[g_{n-2}\left(-\frac{1}{2}\right)\right]^{q-1},
 \end{align}
  \begin{align}
    g_{n-1}\left(\pm\frac{1}{2}\right)&= \sum_{\{S_{2}\}}\exp\left[\beta\left(\pm\frac{J}{2}{S_{2}}+h S_{2}\right)\right]\left[g_{n-2}(S_{2})\right]^{q-1}\nonumber\\
  &= \exp{\left[\beta\left(\pm\frac{J}{4}+\frac{h}{2}\right)\right]}\left[g_{n-2}\left(\frac{1}{2}\right)\right]^{q-1}
 + \exp{\left[\beta\left(\mp\frac{J}{4}-\frac{h}{2}\right)\right]}\left[g_{n-2}\left(-\frac{1}{2}\right)\right]^{q-1}.
 \end{align}
 After calculating all the $g_{n}(S_{0})$ and $g_{n-1}(\sigma_{1})$, we can define the recursion relations for the spin-$\frac{1}{2}$ as:
\begin{equation}
  Z_{n}= \frac{g_{n}\left(\frac{1}{2}\right)}{g_{n}\left(-\frac{1}{2}\right)}\,, \nonumber
\end{equation}
and for the spin-$\frac{5}{2}$ as:
\begin{eqnarray}
 A_{n-1}= \frac{g_{n-1}\left(\frac{5}{2}\right)}{g_{n-1}\left(-\frac{1}{2}\right)}\,, \qquad B_{n-1}= \frac{g_{n-1}\left(-\frac{5}{2}\right)}{g_{n-1}\left(-\frac{1}{2}\right)}\,, \qquad
  C_{n-1}= \frac{g_{n-1}\left(\frac{3}{2}\right)}{g_{n-1}\left(-\frac{1}{2}\right)}\,, \nonumber\\
  D_{n-1}= \frac{g_{n-1}\left(-\frac{3}{2}\right)}{g_{n-1}\left(-\frac{1}{2}\right)}\,, \qquad   E_{n-1}= \frac{g_{n-1}\left(\frac{1}{2}\right)}{g_{n-1}\left(-\frac{1}{2}\right)}\,. \qquad\qquad\qquad
\end{eqnarray}
To investigate our model, we define two order-parameters, the magnetization $M$ and the corresponding quadrupolar moment $Q$. For the
sublattice A, the sublattice magnetization $M_{\text{A}}$ is defined by:
\begin{eqnarray}
 M_{\text{A}} = Z_{\text{A}}^{-1} \sum_{\{S_{0}\}}S_{0}{\exp{(\beta h S_{0})} g_{n}^{q}(S_0)}.
 \end{eqnarray}
After some mathematical manipulations, the sublattice magnetization $M_{\text{A}}$ is explicitly given by:
\begin{eqnarray}
 M_{\text{A}} =  \frac{\exp{\left(\frac{\beta h}{2}\right)}Z_{n}^{q}-\exp{\left(-\frac{\beta h}{2}\right)}}{2\left[\exp{\left(\frac{\beta h}{2}\right)}Z_{n}^{q}+\exp{\left(-\frac{\beta h}{2}\right)}\right]}\,.
\end{eqnarray}
In the same way, we also calculate the two order-parameters for the sublattice B as
follows:
\[M_{\text{B}}=\frac{M_{\text{B}}'}{M_{\text{B}}^{0}}\,, \qquad
 Q_{\text{B}}=\frac{Q_{\text{B}}'}{Q_{\text{B}}^{0}}\,, \]
where:
\begin{align}
 M_{\text{B}}' &= 5\exp{\left(\frac{25}{4}\beta D\right)}\left[\exp{\left(\frac{5}{2}\beta h\right)}A_{n-1}^{q}-\exp{\left(-\frac{5}{2}\beta h\right)}B_{n-1}^{q}\right]
 + 3\exp{\left(\frac{9}{4}\beta D\right)}\left[\exp{\left(\frac{3}{2}\beta h\right)}C_{n-1}^{q}\right.\nonumber\\ & \quad\left.-\exp{\left(-\frac{3}{2}\beta h\right)}D_{n-1}^{q}\right]
  +\exp{\left(\frac{1}{4}\beta D\right)}\left[\exp{\left(\frac{1}{2}\beta h\right)}E_{n-1}^{q}-\exp{\left(-\frac{1}{2}\beta h\right)}\right],
\end{align}
\begin{align}
M_{\text{B}}^{0} &= 2\exp{\left(\frac{25}{4}\beta D\right)}\left[\exp{\left(\frac{5}{2}\beta h\right)}A_{n-1}^{q} +\exp{\left(-\frac{5}{2}\beta h\right)}B_{n-1}^{q}\right]
+ 2\exp{\left(\frac{9}{4}\beta D\right)}\left[\exp{\left(\frac{3}{2}\beta h\right)}C_{n-1}^{q}\right.\nonumber\\ & \quad\left.+\exp{\left(-\frac{3}{2}\beta h\right)}D_{n-1}^{q}\right]
 +2\exp{\left(\frac{1}{4}\beta D\right)}\left[\exp{\left(\frac{1}{2}\beta h\right)}E_{n-1}^{q}+\exp{\left(-\frac{1}{2}\beta h\right)}\right],
\end{align}
\begin{align}
Q_{\text{B}}' &= 25\exp{\left(\frac{25}{4}\beta D\right)}\left[\exp{\left(\frac{5}{2}\beta h\right)}A_{n-1}^{q} +\exp{\left(-\frac{5}{2}\beta h\right)}B_{n-1}^{q}\right]
+ 9\exp{\left(\frac{9}{4}\beta D\right)}\left[\exp{\left(\frac{3}{2}\beta h\right)}C_{n-1}^{q}\right.\nonumber\\ & \quad\left.+\exp{\left(-\frac{3}{2}\beta h\right)}D_{n-1}^{q}\right]
+\exp{\left(\frac{1}{4}\beta D\right)}\left[\exp{\left(\frac{1}{2}\beta h\right)}E_{n-1}^{q}+\exp{\left(-\frac{1}{2}\beta h\right)}\right],
\end{align}
\begin{align}
Q_{\text{B}}^{0} &= 4\exp{\left(\frac{25}{4}\beta D\right)}\left[\exp{\left(\frac{5}{2}\beta h\right)}A_{n-1}^{q}+\exp{\left(-\frac{5}{2}\beta h\right)}B_{n-1}^{q}\right]
+ 4\exp{\left(\frac{9}{4}\beta D\right)}\left[\exp{\left(\frac{3}{2}\beta h\right)}C_{n-1}^{q}\right.\nonumber\\ & \quad\left.+\exp{\left(-\frac{3}{2}\beta h\right)}D_{n-1}^{q}\right]
+4\exp{\left(\frac{1}{4}\beta D\right)}\left[\exp{\left(\frac{1}{2}\beta h\right)}E_{n-1}^{q}+\exp{\left(-\frac{1}{2}\beta h\right)}\right].
\end{align}

In order to determine the compensation temperature, one has to define the global magnetization $M_{T}$ of the model
which is given by:
\begin{eqnarray}
 M_{T}=\frac{M_\text{A}+M_\text{B}}{2}\,.
\end{eqnarray}

To really study the model in detail and  single out the effect of the crystal-field and of the applied magnetic field on the magnetic properties
of the model, we have also examined the thermal variations of the response functions, i.e., the susceptibilities, the specific heat and the internal
energy defined respectively by:
\begin{eqnarray}
 \chi_{\text{Total}} = \chi_\text{A} + \chi_\text{B} = \left(\frac{\partial{M_\text{A}}}{\partial{h}} \right)_{h=0} + \left(\frac{\partial{M_\text{A}}}{\partial{h}} \right)_{h=0}\,,
\end{eqnarray}
\begin{eqnarray}
  C = -{\beta}^{2}\frac{{\partial}^{2}{(-\beta F')}}{\partial{{\beta}^{2}}}\,,
\end{eqnarray}
\begin{eqnarray}
 \frac{U}{N|J|}=-k_\text{B}T^{2}\frac{\partial}{\partial T}\left(\frac{F'}{k_\text{B}T}\right),
\end{eqnarray}
where $ F' $ is the free energy of the model.
\section{Definition of the critical temperature}

The Curie temperature or the second-order transition temperature $T_\text{c}$ is the temperature at which both
 sublattice magnetizations
 and the global magnetization continuously go to zero. $T_\text{c}$ separates the ordered ferrimagnetic phase (F) from the disordered
 paramagnetic phase (P).
 At $T_\text{c}\,$, one can obtain explicit expressions of the recursion relations
 as follows:\\
  for the spin-$\frac{1}{2}\,$,
   \begin{equation}
Z_{n}= 1,
\end{equation}
and for the spin-$\frac{5}{2}\,$,
\begin{equation}
A_{n-1} =B_{n-1} =\frac{\cosh\left(\frac{5\beta J}{4}\right)}{\cosh\left(\frac{\beta J}{4}\right)}\,, \qquad
C_{n-1} =D_{n-1} =\frac{\cosh\left(\frac{3\beta J}{4}\right)}{\cosh\left(\frac{\beta J}{4}\right)}\,,\qquad
 E_{n-1}=1.
\end{equation}
 In addition to the thermal variations of the order-parameters and the global magnetization of the model,
 we also calculate and analyze the free energy $F'$ of the model in order to identify
 the first-order transition temperature $T_\text{t}$. Thus, using the definition of the free energy $ F' = -k_{\textrm{B}}T\ln(Z)$ in the thermodynamic limit
 $(n\rightarrow\infty)$ and in order to introduce the recursion relations, we can rewrite the free energy as:
 \begin{eqnarray}
  F'/J=-\frac{1}{{\beta}'}\left[ \frac{q-1}{2-q} \ln{F_1}+\frac{1}{2-q}\ln{F_2}+\ln{F_3}\right],
 \end{eqnarray}
where ${\beta}'=\beta J$, $F_1=g_{n-1}(-1/2)/g_{n}^{q-1}(-1/2)$,\quad $F_2=g_{n}(-1/2)/g_{n-1}^{q-1}(-1/2)$
 and  $F_3= Z/g_{n}^{q}(-1/2)$.

After some mathematical manipulations, the free energy expression in terms of recursion relations is
explicitly given by:
\begin{align}
 F'/J &= -\frac{1}{{\beta}'}\left(\frac{q-1}{2-q}\ln\left\{\exp{\left[\beta\left(-\frac{J}{4} +\frac{h}{2}\right)\right]}Z_{n}^{q-1} +\exp{\left[\beta\left(\frac{J}{4}-\frac{h}{2}\right)\right]}\right\}\right)\nonumber\\
 &\quad -\frac{1}{{\beta}'}\left\{\ln\left[\exp{\left(\beta\frac{h}{2}\right)}Z_{n}^{q} +\exp{\left(-\beta\frac{h}{2}\right)}\right]\right\}\nonumber\\
 &\quad -\frac{1}{{\beta}'}\left(\frac{1}{2-q}\ln\left\{\exp{\left[\beta\left(-\frac{5J}{4} +\frac{25 D}{4} +\frac{5h}{2}\right)\right]}A_{n-1}^{q-1} +\exp{\left[\beta\left(\frac{5J}{4} +\frac{25D}{4} -\frac{5h}{2}\right)\right]}B_{n-1}^{q-1} \right.\right.\nonumber \\
 &\quad +\exp{\left[\beta\left(-\frac{3J}{4} +\frac{9D}{4} +\frac{3h}{2} \right)\right]}C_{n-1}^{q-1} +\exp{\left[\beta\left(\frac{3J}{4} +\frac{9D}{4} -\frac{3h}{2} \right)\right]}D_{n-1}^{q-1}\nonumber\\
 &\quad \left.\left.+\exp{\left[\beta\left(-\frac{J}{4} +\frac{D}{4} +\frac{h}{2}\right)\right]}E_{n-1}^{q-1} +\exp{\left[\beta\left(\frac{J}{4} +\frac{D}{4} -\frac{h}{2}\right)\right]}\right\}\right).
\end{align}

 We also investigate the compensation temperature $T_{\text{comp}}$ at which the global magnetization vanishes
 while both sublattice magnetizations cancel each other.
$T_{\text{comp}}$ is found by locating the crossing point between the absolute values of sublattice magnetizations, i.e.,
   \begin{equation}
    |M_{\text{A}}(T_{\text{comp}})|= |M_{\text{B}}(T_{\text{comp}})|.
   \end{equation}
  Considering different definitions of the critical temperature, we can now investigate the
  thermal variations of the calculated thermodynamical quantities of interest and display the temperature phase diagrams
  of the model for $q=3, 4, 5$ and $6$.
  \section{Numerical results and discussions}\label{sec4}

In this section, we present and discuss the results we obtained for
  the thermal variations of the order-parameters, the response functions, the internal energy and the temperature
  phase diagrams of the model. We begin discussions with the ground-state
  phase diagram which is necessary for understanding the obtained temperature phase diagrams.
  \subsection{Ground-state phase diagram}
\begin{figure}[!b]
\begin{center}
\includegraphics[angle=0,width=0.55\textwidth]{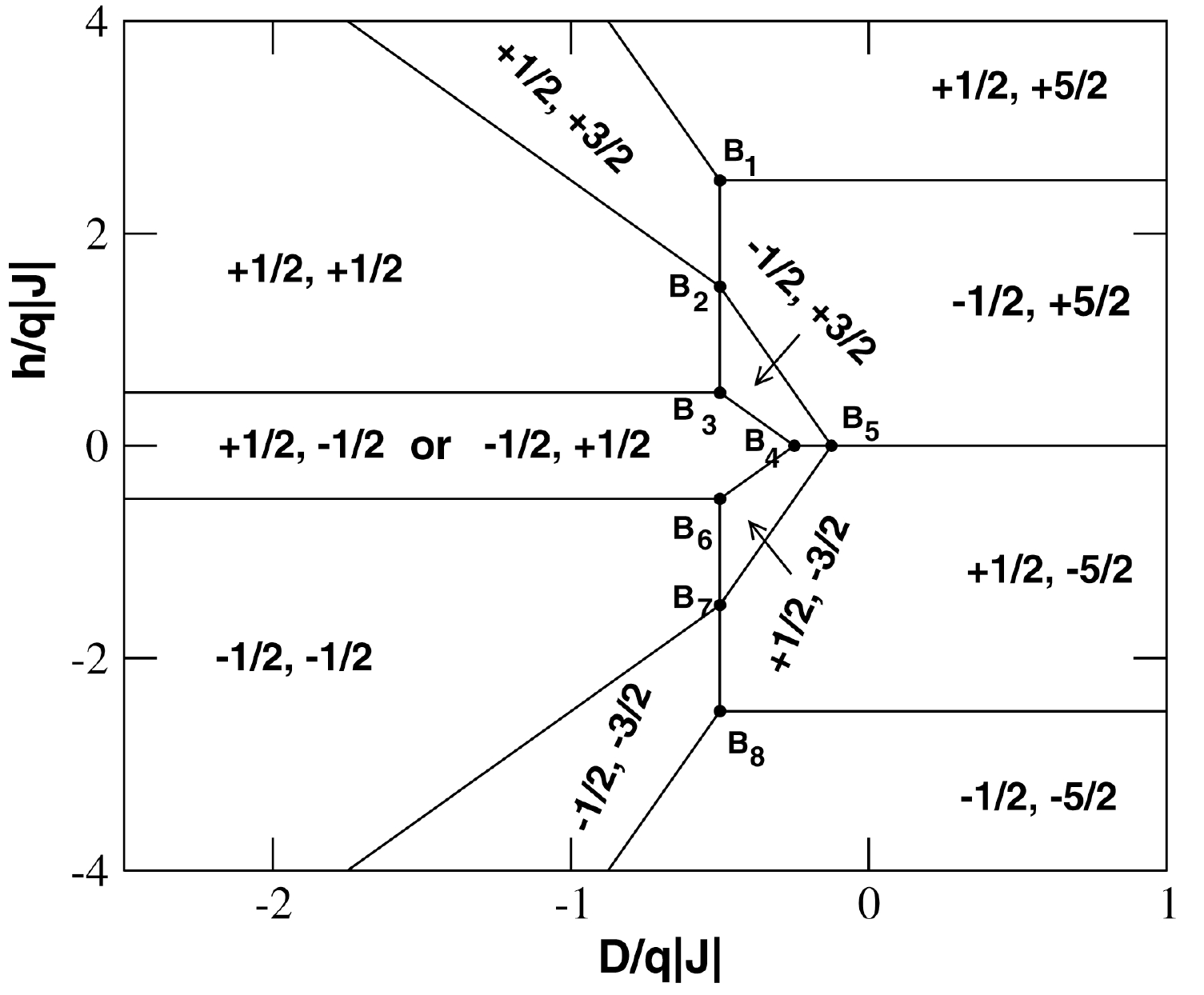}
\end{center}
\caption{Ground-state phase diagram of the mixed spin-$\frac{1}{2}$ and spin-$\frac{5}{2}$ Ising ferrimagnetic model with the same
crystal-field $D$ for the two sublattices in the $(h/q|J|, \, D/q|J|)$ plane. There exist eleven stable phases.
Along the $D/q|J|$-axis and for all values of $q$, two hybrid phases may appear at the multicritical
points $B_4$ and $B_5$.}
\label{fig2}
\end{figure}

   Before presenting the numerical results for the temperature dependence of magnetic
   properties of the model, we first investigate the ground-state phase diagram.
   The ground-state structure of the model can be represented by comparing the values of the energy $H_0$
    for different spin configurations which can be expressed as:
  \begin{eqnarray}
   H_0=S\sigma-\frac{1}{q |J|}\left[D  {\sigma}^{2}+h (S+\sigma)\right].
  \end{eqnarray}
 We only get eleven possible pairs of spins.
Computational calculations of the corresponding energies in the $(h/q|J|,\, D/q|J|)$ plane yields the ground-state phase diagram displayed in figure~\ref{fig2}.
This diagram shows some interesting
features of the model, in particular, the existence of eight multicritical points $(B_1, B_2,\cdots , B_{8})$ and coexistence lines
where the spin pair energy of some phases is the same. In the absence of the magnetic field, for all values of $q$ and $D/q|J|$,
$M_\text{B}$ shows five saturation values whereas for $M_\text{A}$, $\pm{\frac{1}{2}}$ is the only saturation value. Thus, we get the ferrimagnetic phases
$\text{ F}(\pm{\frac{1}{2}}, \mp{\frac{5}{2}})$, $\text{ F}(\pm{\frac{1}{2}}, \mp{\frac{3}{2}}) $, $\text{ F}(\pm{\frac{1}{2}}, \mp{\frac{1}{2}}) $ and
at the borders of these phases, two hybrid phases:
$\text{ F}(\pm{\frac{1}{2}}, \mp{1})$, $\text{ F}(\pm{\frac{1}{2}}, \mp{2})$ at the multicritical points $B_4$ and $B_5$, respectively. These hybrid phases should correspond to cases
where the sublattice B is half-half covered by spins of the two neighboring phases. It is important to indicate that
 the ground-state phase diagram is very useful because it helps to check the reliability of the theoretical results and to classify different phase domains of the model for
 the temperature dependence phase diagrams.

\subsection{Thermal variations of the order-parameters, the response functions and the internal
energy}

\begin{figure}[!b]
\begin{center}
\includegraphics[angle=0,width=0.47\textwidth]{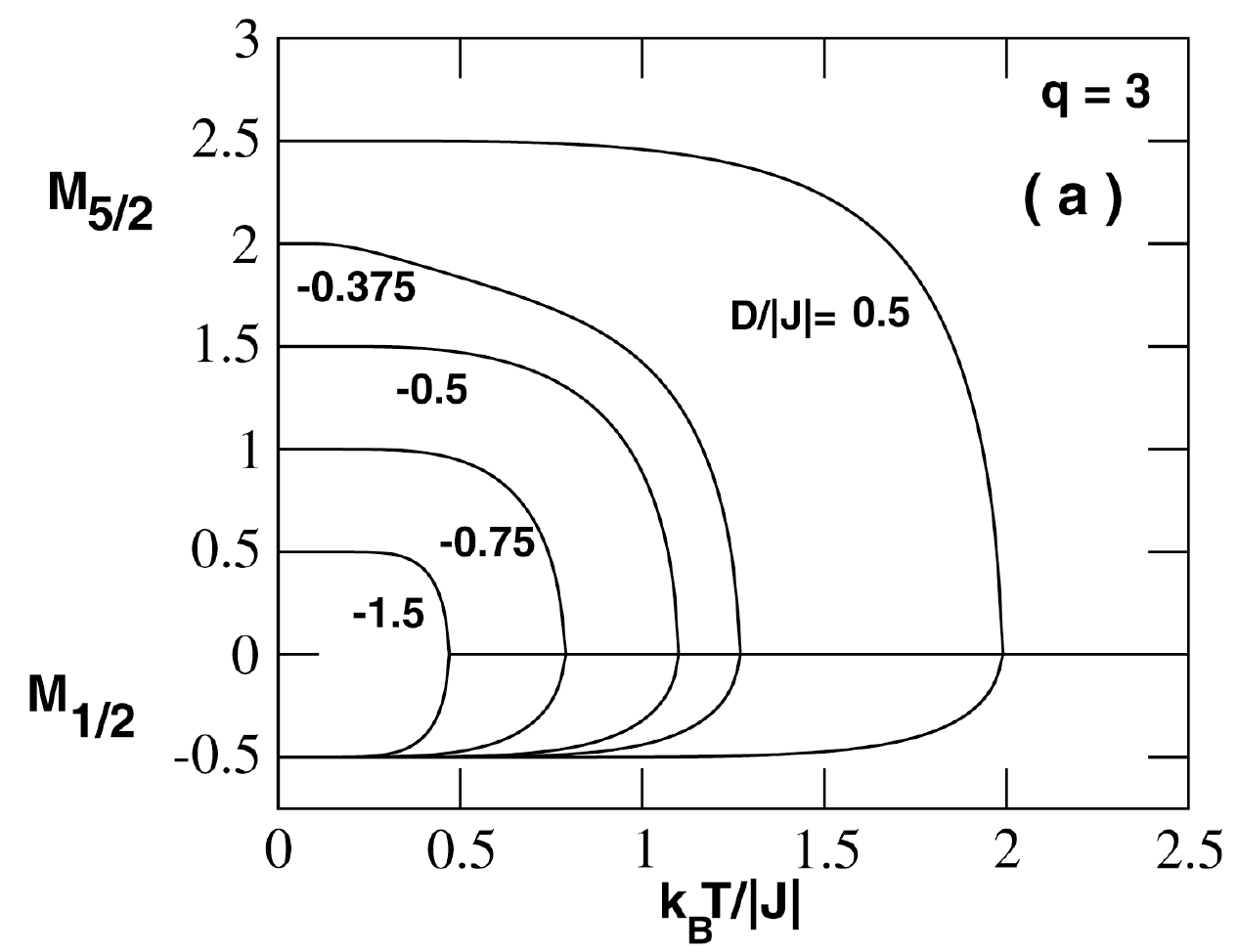} \hspace{0.5cm}
\includegraphics[angle=0,width=0.47\textwidth]{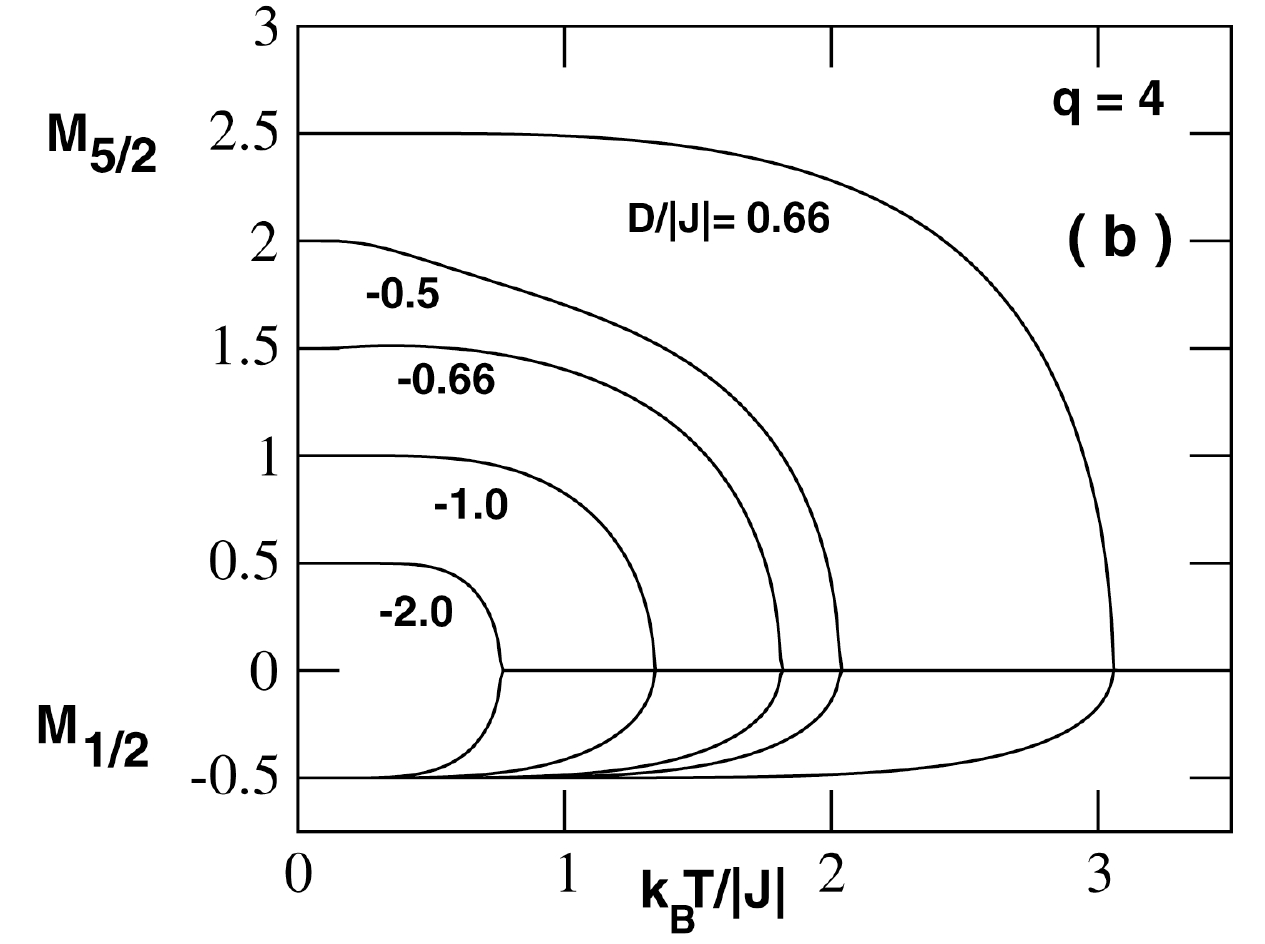}\\
\vspace{.2cm}
\includegraphics[angle=0,width=0.47\textwidth]{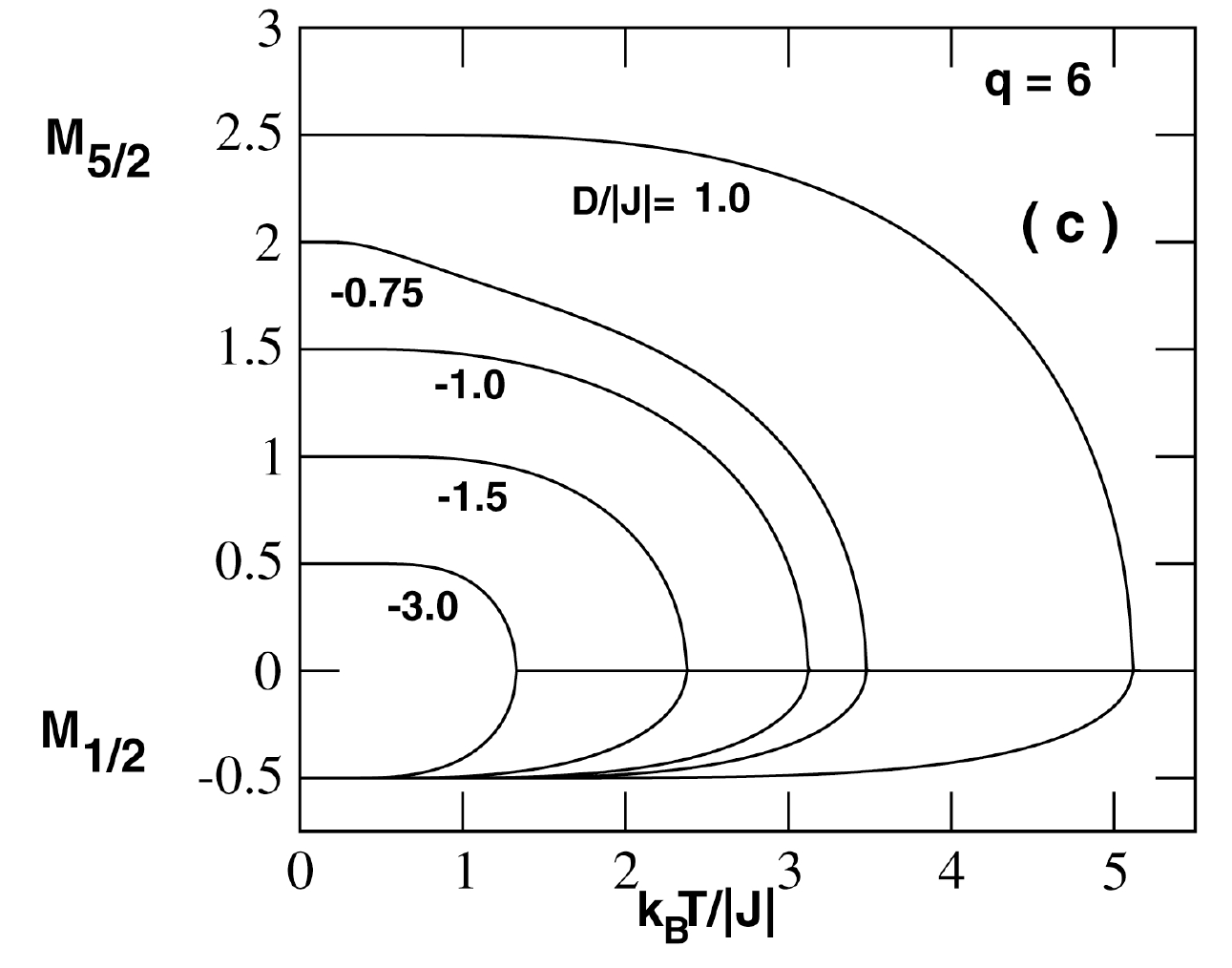}
\end{center}
\caption{
  Sublattice magnetizations of the model as functions of the reduced temperature $k_{\textrm{B}}T/|J|$ when $q=3, 4$ and $6$ for
  various values of the crystal-field interactions $D/|J|$. Panel (a): Curves are displayed
  for $q=3$ and selected values of $D/|J|$ indicated on the curves. Panel (b): curves are
  displayed for $q=4$ and selected values of $D/|J|$ indicated on the curves. Panel (c): curves
  are displayed for $q=6$ and selected values of $D/|J|$ indicated on the curves. For all values of $q$,
  the sublattice magnetization curves are all continuous. }
\label{fig3}
\end{figure}

As it is explained above, thermal variations of the order-parameters, the response
 functions and the internal energy
 for the present model were calculated in terms of recursion relations. Thermal
 variations of the order-parameters are crucial in obtaining the temperature dependence phase
 diagrams of the model. Thus, when the magnetization curves go to zero continuously
 separating the ferrimagnetic phase from the paramagnetic phase, one gets the second-order phase
 transition or Curie temperature, i.e., the temperature at which magnetizations become zero.
 In the case of a jump in the magnetization curves followed by a discontinuity of the first derivative
 of the free-energy $F'$, one gets a first-order transition temperature.
 Besides these two temperatures, there is another temperature referred to as
 compensation temperature defined as the temperature where the global
 magnetization becomes zero prior to the critical temperature. Therefore, in order to
 identify transitions and compensation lines, one should
 study the thermal behaviors of the considered thermodynamical quantities of the model. Now, we  present
 some results on
 the thermal behaviors of the order-parameters, the response functions and the internal energy in the
 the absence of the magnetic field $h$ when $q=3, 4$ and $6$.

\begin{figure}[!b]
\vspace{-2mm}
\begin{center}
\includegraphics[angle=0,width=0.4\textwidth]{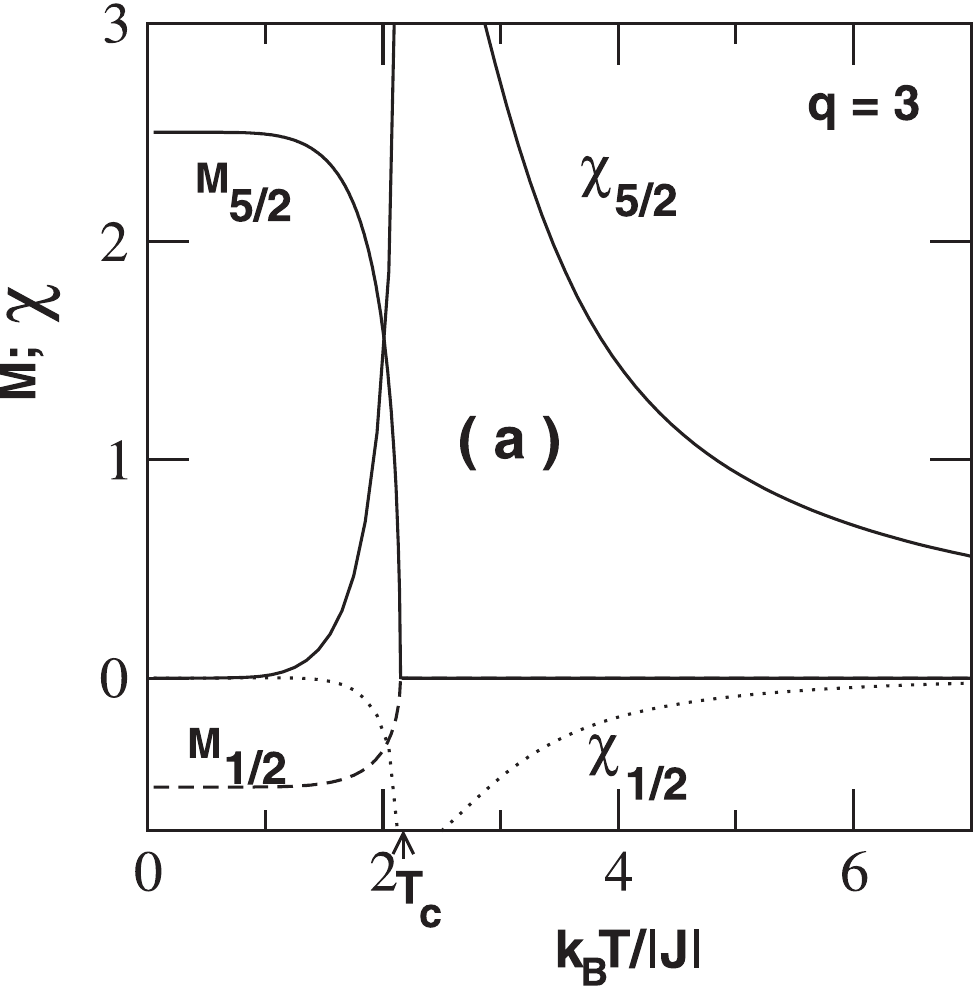} \hspace{1.3cm}
\includegraphics[angle=0,width=0.4\textwidth]{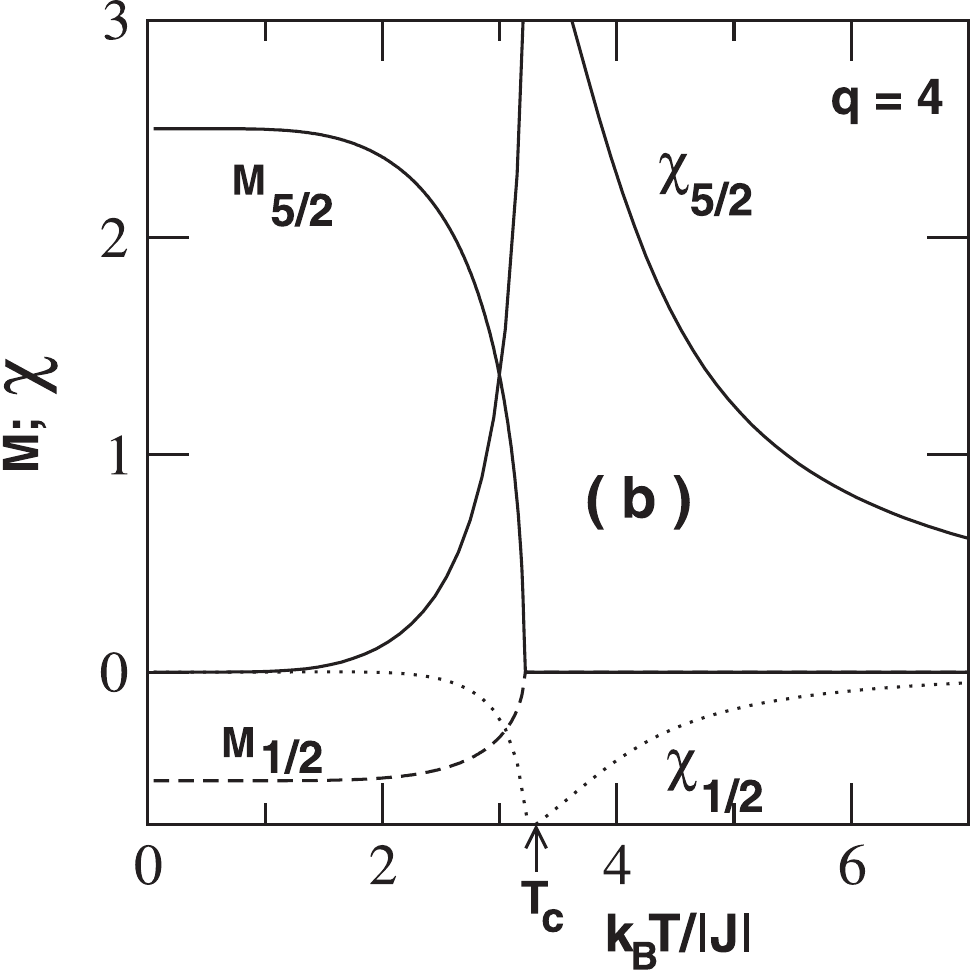}\\
\vspace{0.2cm}
\includegraphics[angle=0,width=0.4\textwidth]{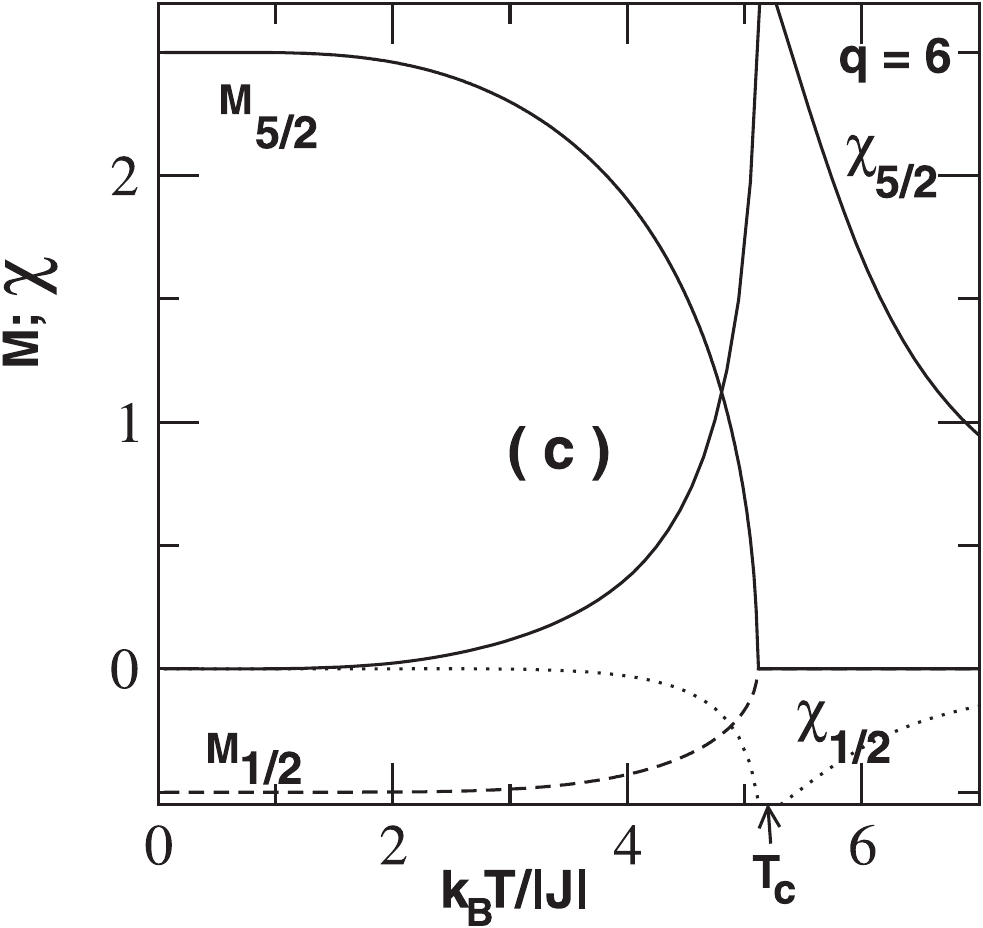}
\end{center}
\caption{
Thermal variations of sublattice magnetizations and corresponding susceptibilities
are calculated for $q=3, 4, 6$ and the reduced crystal-field $ D/|J|=1 $ as shown
from panel (a) to panel (c). Values of the physical parameters
considered for the system are indicated in different panels.  $T_\text{c}$  indicates the second-order temperature.}
\label{fig4}
\end{figure}

Figure~\ref{fig3} displays some thermal variations of the sublattice magnetizations
  $M_{1/2}$ and $M_{5/2}$ when $q=3, 4$ and $6$ for selected values of the crystal-field $D/|J|$.
  From panels (a) to (c), we have depicted the thermal behaviors of
  sublattice magnetizations $M_{1/2}$ and $M_{5/2}$ as functions of the temperature for selected
  values of $D/|J|$ when $q=3, 4$ and 6. The results are in perfect agreement with the
  ground-state phase diagram concerning the saturation values. Indeed, $M_{1/2}$ increases from
  its unique saturation value $\pm{\frac{1}{2}}$ with increasing temperature whereas $M_{5/2}$ shows
  five saturation values. The behaviors of the sublattice
  magnetizations $M_{1/2}$ and $M_{5/2}$ are quite similar. We notice that all the curves are
  continuous and the Curie temperature $T_\text{c}$ at which both magnetizations curves go to zero increases with
  the strength of the crystal-field $D/|J|$ and the coordination number $q$. Moreover, by comparing figure~\ref{fig3}
  to figure~\ref{fig4} of \cite{ra18}, the sublattice magnetizations show similar thermal variations.

  To explain in detail the results obtained in figure~\ref{fig3}, we have investigated the thermal
  variations of order-parameters, the corresponding response functions and the internal energy.

  Thus, on the one hand, as shown in figure~\ref{fig4}, we have displayed the thermal behaviors of the sublattice
  magnetizations and corresponding susceptibilities when $q=3, 4, 6$ and $D/|J|=1$. In this figure,
  one can notice that the model only exhibits the second-order phase transition, and the transition temperature~$T_\text{c}$
  at which the transition occurs increases with an increasing coordination number $q$. Here, $T_\text{c}$ separates the
  ferrimagnetic phase $\text{F}(\pm{\frac{1}{2}}, \mp{\frac{5}{2}})$ from the paramagnetic phase (P) and $T_\text{c}/|J|=2.1724$
  (respectively $T_\text{c}/|J|=3.2575$ and $5.1398$)
  for $q=3$ (respectively  for $q=4$ and $6$). Also, one remarks that for
  $T\rightarrow T_\text{c}\,$, $\chi_{1/2} \rightarrow -\infty$ whereas $\chi_{5/2} \rightarrow +\infty$.
   For $T> T_\text{c}\,$, the susceptibility $\chi_{1/2}$ rapidly increases whereas the susceptibility $\chi_{5/2}$
   rapidly decreases when the temperature increases and is very far from the Curie temperature~$T_\text{c}\,$,
  $\chi_{1/2} \rightarrow 0$ and $\chi_{5/2} \rightarrow 0$.

\begin{figure}[!b]
\begin{center}
\includegraphics[angle=0,width=0.85\textwidth]{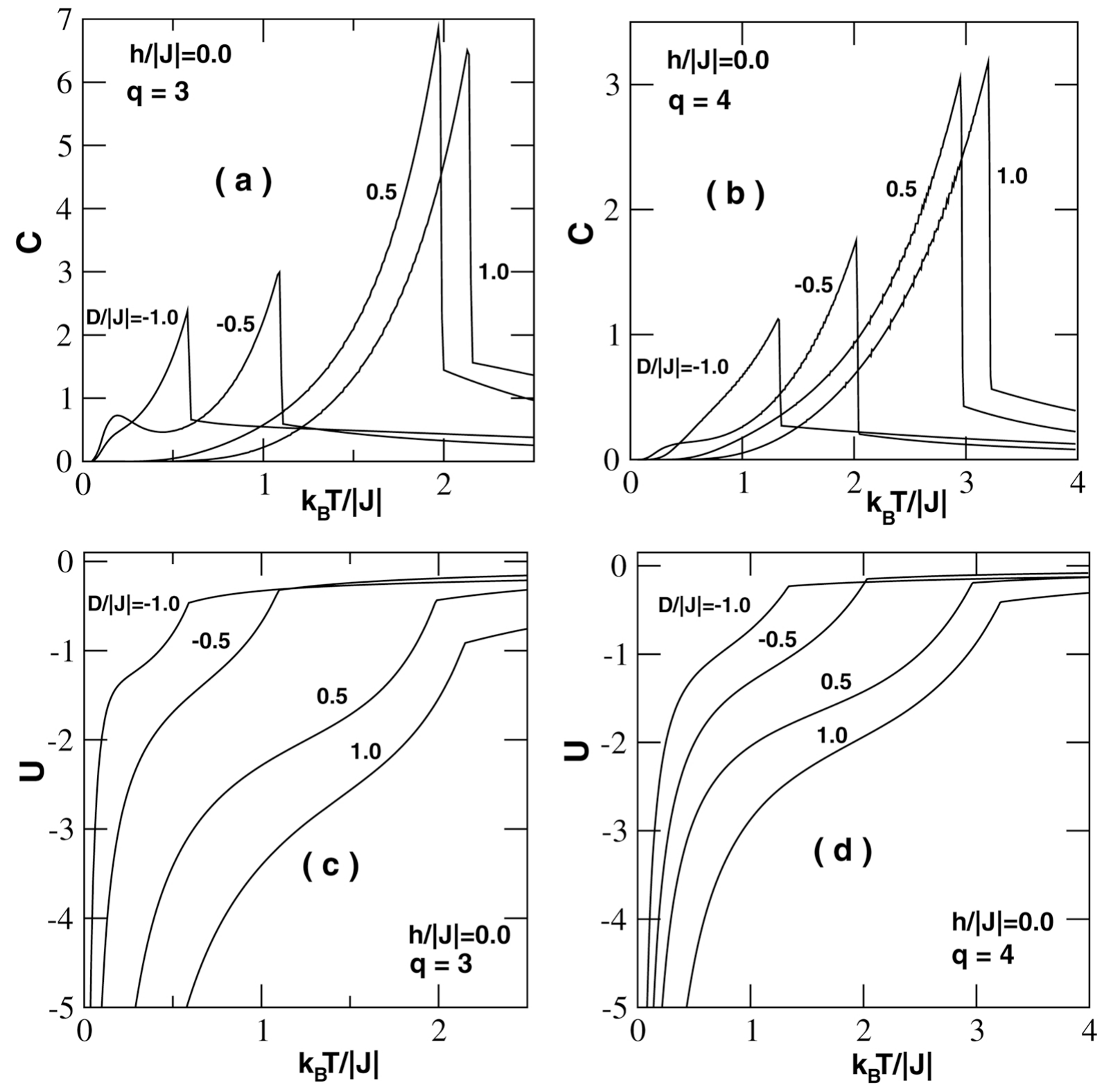}
\end{center}
\caption{
Thermal variations of the specific heat and internal energy
are calculated for $q=3, 4$ and selected values of the crystal-field $ D/|J| $ as shown in the
figures from panel (a) to panel (d). Values of the physical parameters
considered for the system are indicated in different panels. Analysis of different panels of
this figure shows that the model only exhibits a second-order transition.}
\label{fig5}
\end{figure}

  On the other hand, to really confirm that the model only exhibits the second-order transition for all values of $q$,
  we have plotted  in figure~\ref{fig5} the temperature dependence variations of the specific heat and the
  internal energy for various values of the crystal-field as indicated in the figure. Both the specific heat and
  the internal energy rapidly increase with an increasing temperature and  make a peak without jump discontinuities at
  the same $T_\text{c}\,$. By increasing the strength of the crystal-field and the coordination number,
  the $ T_\text{c}$ at which the transition occurs, increases and this is easily
  observed by comparing the results from different panels of figure~\ref{fig5}. The results obtained in this figure also confirm that
  the model only presents second-order transition for all values of the coordination number~$q$.

  Let us now discuss the thermal variations of sublattice magnetizations, the corresponding response functions
  and the internal energy of the system in the presence of a longitudinal magnetic field~$h$.

\begin{figure}[!b]
\begin{center}
\includegraphics[angle=0,width=0.4\textwidth]{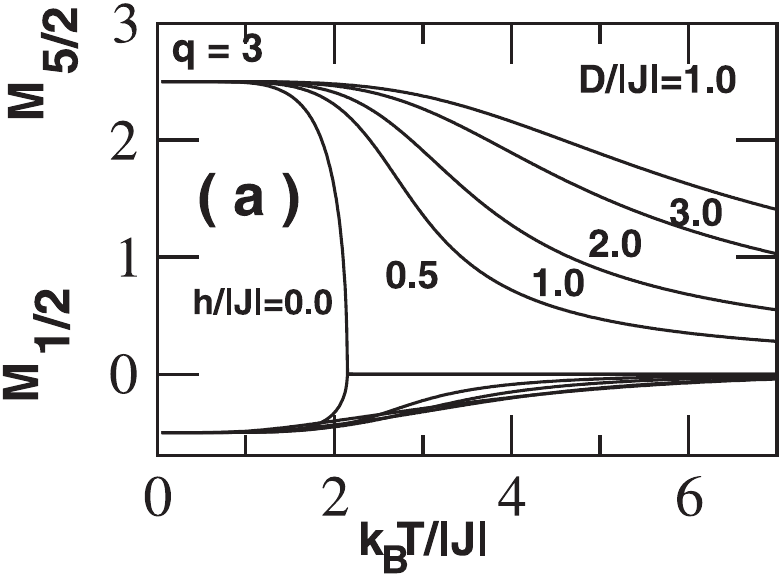} \hspace{1.3cm}
\includegraphics[angle=0,width=0.4\textwidth]{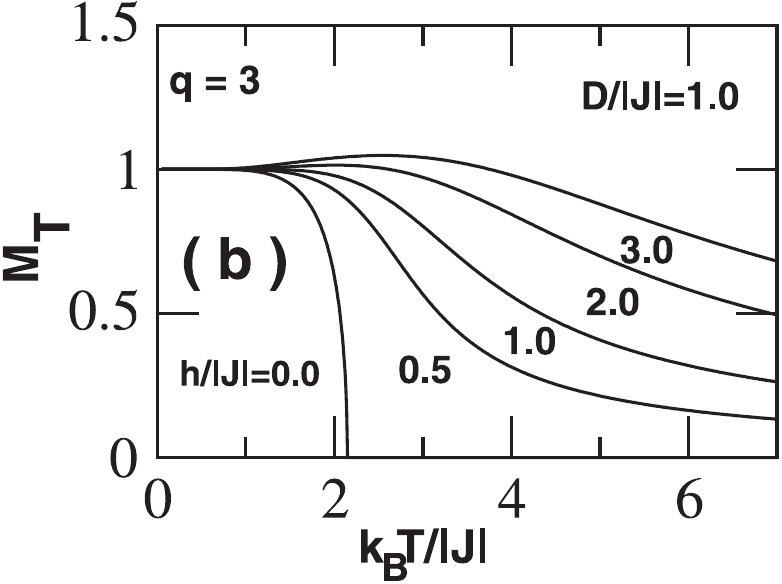}\\
\vspace{0.2cm}
\includegraphics[angle=0,width=0.4\textwidth]{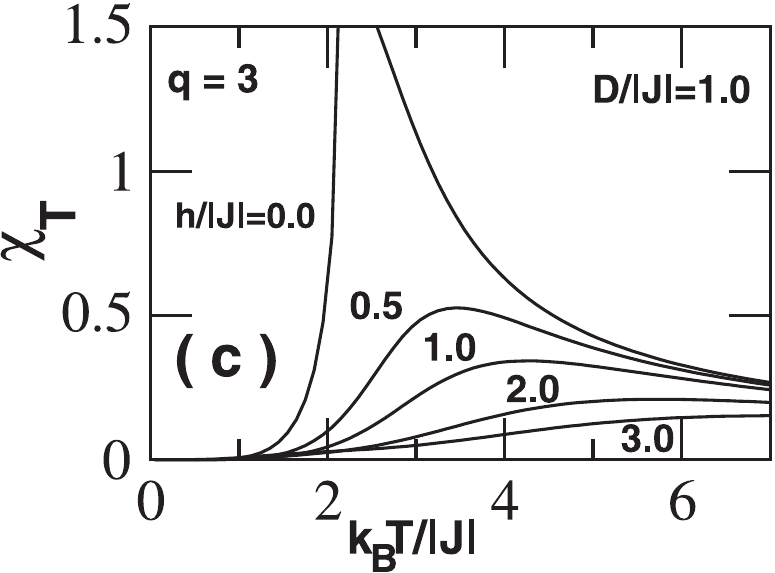} \hspace{1.3cm}
\includegraphics[angle=0,width=0.4\textwidth]{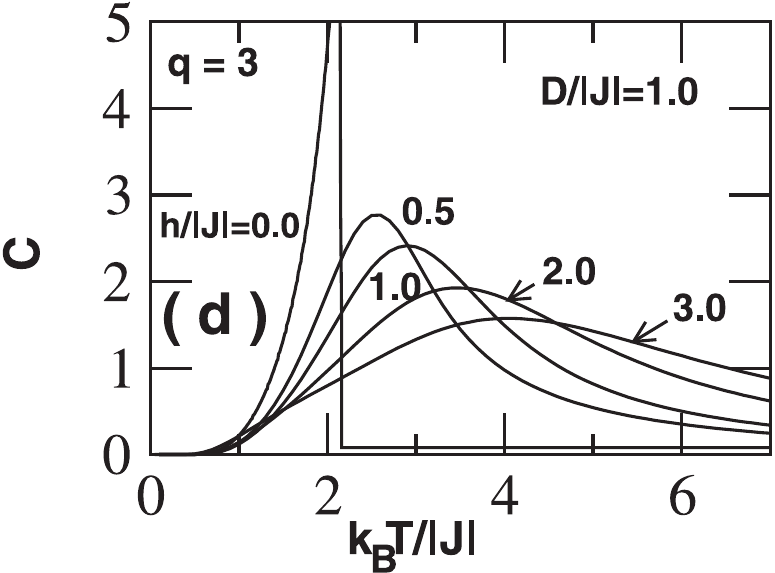}\\
\vspace{0.2cm}
\includegraphics[angle=0,width=0.4\textwidth]{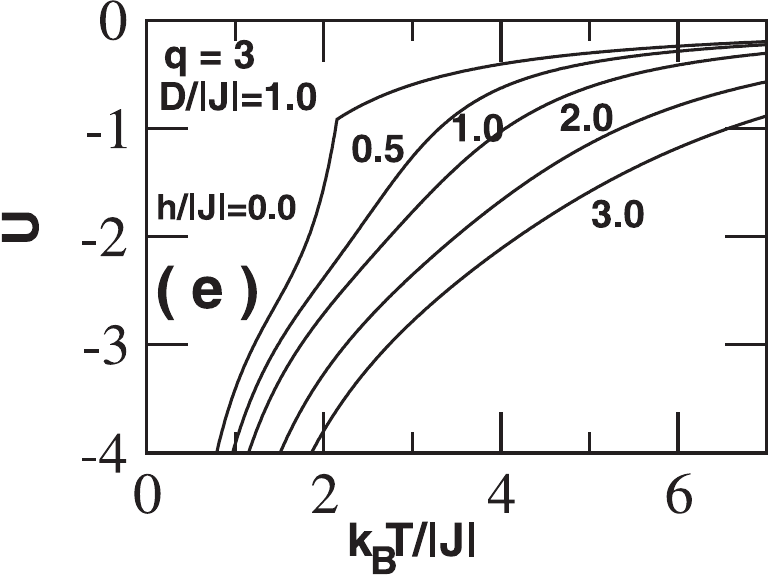}
\end{center}
\caption{
 Thermal behaviors of the sublattice magnetizations, the global magnetization, the corresponding response functions and the internal
 energy of the model when $D/|J|=1.0$ and $q=3$ for selected values of the magnetic field $h/|J|$ as shown
  in different panels. From the analysis of different panels, one can conclude that the model
 only shows temperature phase transition when $h/|J|=0$. For $h/|J| \ne 0$, the remaining magnetizations
 are more important when the value of the magnetic field $h/|J|$ increases.}
\label{fig6}
\end{figure}

  Figure~\ref{fig6} expresses the effects of an applied magnetic field $h$ on the magnetic properties of the model
  when $q=3$ and $D/|J|=1.0$ for selected values of $h/|J|$.
  From panel (a) to panel (b), the sublattice magnetizations and the global magnetization continuously fall from their saturation values
  to non-zero values when the temperature increases. The remaining values of magnetizations are
  more important when the value of the applied magnetic field $h/|J|$ increases. Thus, one can observe that the system does not
  present any transition when $h/|J| \ne 0$. It is important to indicate that in the case of $h/|J|=0$, the model exhibits
  the second-order transition at a Curie temperature $T_{\text{c}}/|J|=2.1724$, where the two sublattice magnetizations and the global magnetization continuously go to
  zero after falling from their saturation values at $T=0$. In panels (c)--(e), we have displayed the temperature
  dependence of the total susceptibility $\chi_{T}\,$, the specific heat $C$ and the internal energy $U$. One can see from these panels that
 the response functions and the internal energy also indicate a second-order transition which occurs at the same
 $T_{\text{c}}/|J|$ as in the case of $h/|J|=0$. For $h/|J| \ne 0$ and $T>T_\text{c}\,$, the response functions exhibit a maximum
 and the height of the maximum decreases when the value of the applied magnetic field increases.

\begin{figure}[!t]
\begin{center}
\includegraphics[angle=0,width=0.85\textwidth]{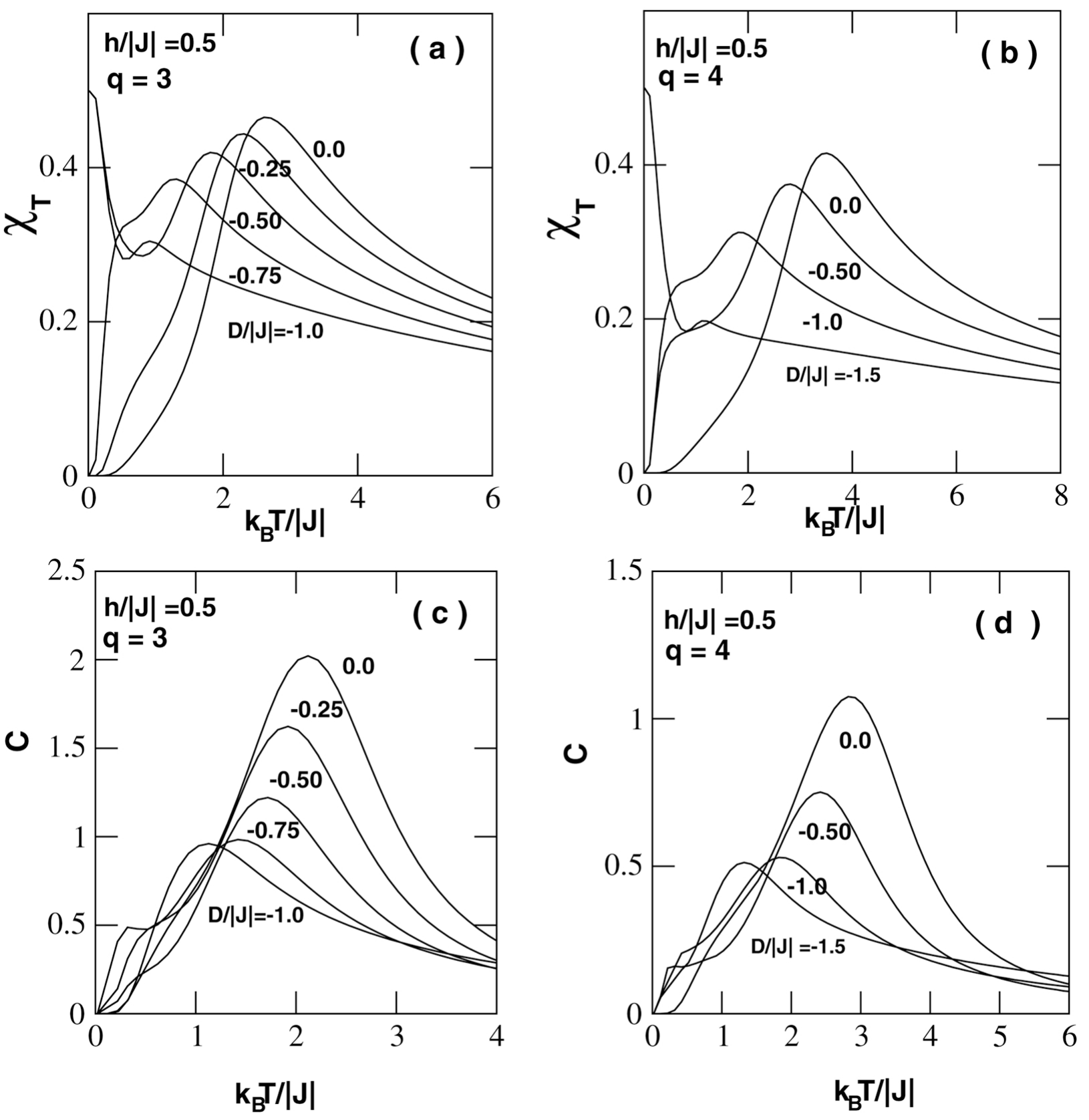}
\end{center}
\vspace{-.2cm}
\caption{
Temperature variations of the response functions of the model for selected values of the crystal-field
$D/|J|$ as indicated on different curves illustrated when $h/|J|=0.5$ and $q=3, 4$.}
\label{fig7}
\end{figure}

 To show the effect of $D/|J|$ on the system properties for $h/|J| \ne 0$, we have illustrated in figure~\ref{fig7} the thermal variations of the
 response functions for some values of the system parameters: $h/|J|=0.5$; $q=3, 4$ and varying $D/|J|$.
 Considering the different panels of figure~\ref{fig7}, one observes that the response functions show interesting
 behaviors. Indeed, the two studied response functions globally show a maximum at a certain value of the temperature.
 This temperature increases with the coordination number and the strength of the crystal-field. It is important
 to mention that the height of the maximum of the two response functions also increases by increasing
 the strength of the crystal-field $D/|J|$ but the opposite holds when the coordination number $q$
 increases.

\begin{figure}[!t]
\begin{center}
\includegraphics[width=0.55\textwidth]{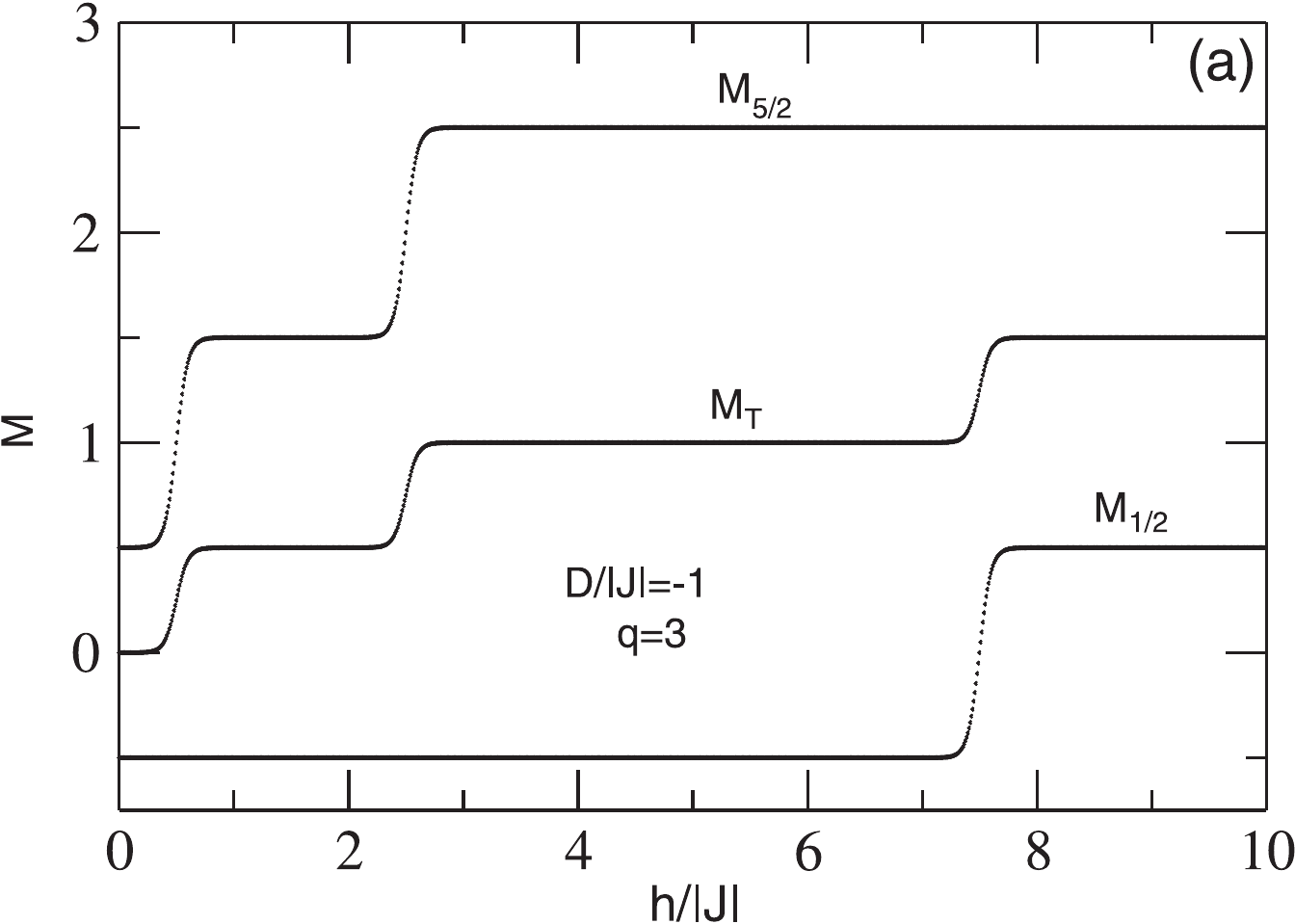}\\ \vspace{.2cm}
\includegraphics[width=0.55\textwidth]{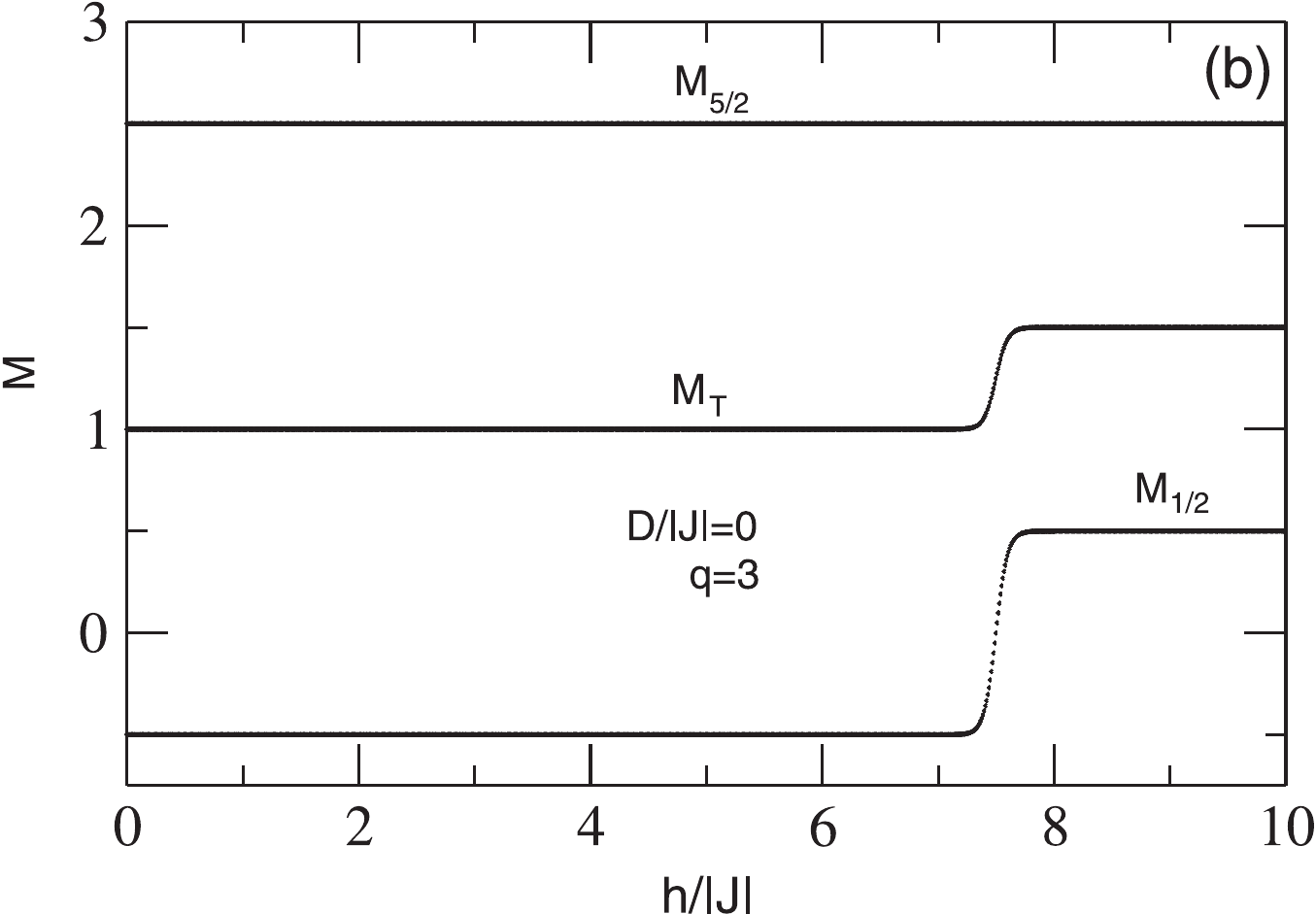}
\end{center}
\caption{
$M_{1/2}$, $M_{5/2}$ and $M_T$ plotted as functions of the magnetic field $h$ for selected values of the crystal-field
when $q=3$ as indicated in different panels.}
\label{fig8}
\end{figure}

\begin{figure}[!t]
\centering
\begin{minipage}{0.33\linewidth}
\begin{center}
\includegraphics[width=\textwidth,height=5cm]{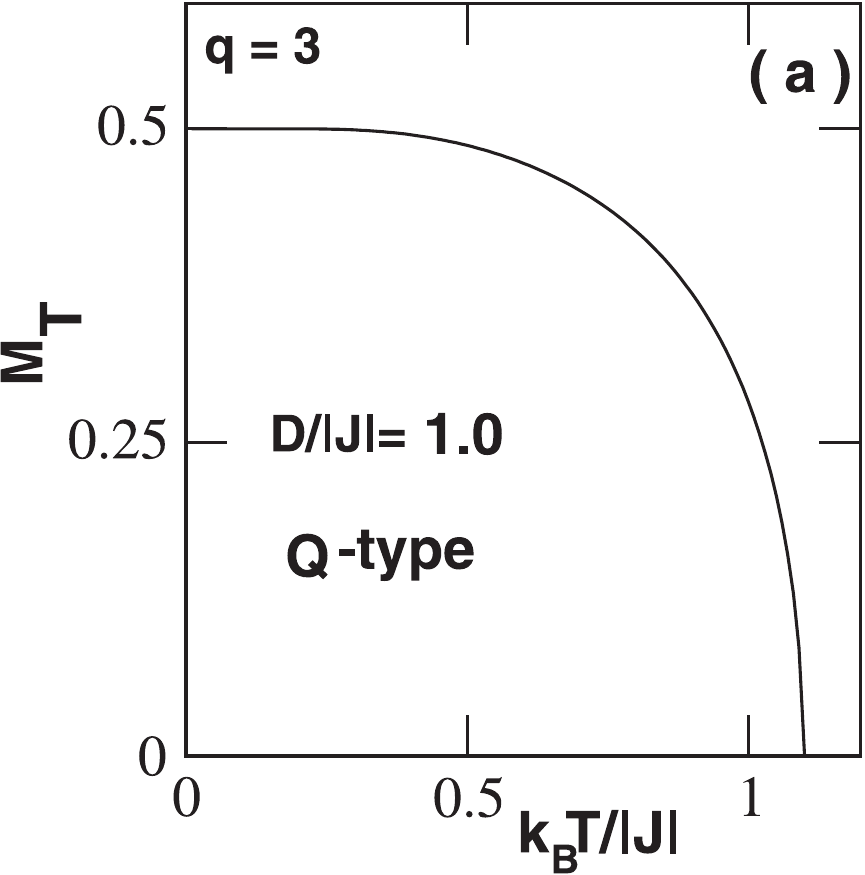}
\end{center}
\end{minipage}
\begin{minipage}{0.33\linewidth}
\begin{center}
\includegraphics[width=\textwidth,height=5cm]{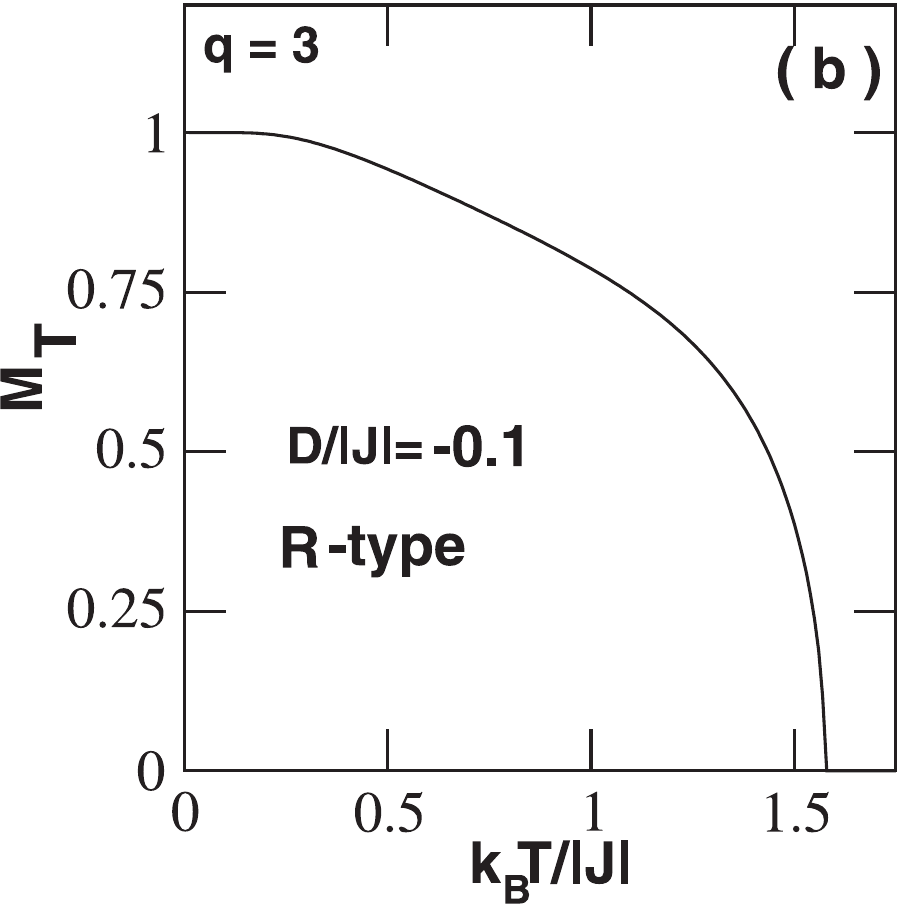}
\end{center}
\end{minipage}
\begin{minipage}{0.33\linewidth}
\begin{center}
\includegraphics[width=\textwidth,height=5cm]{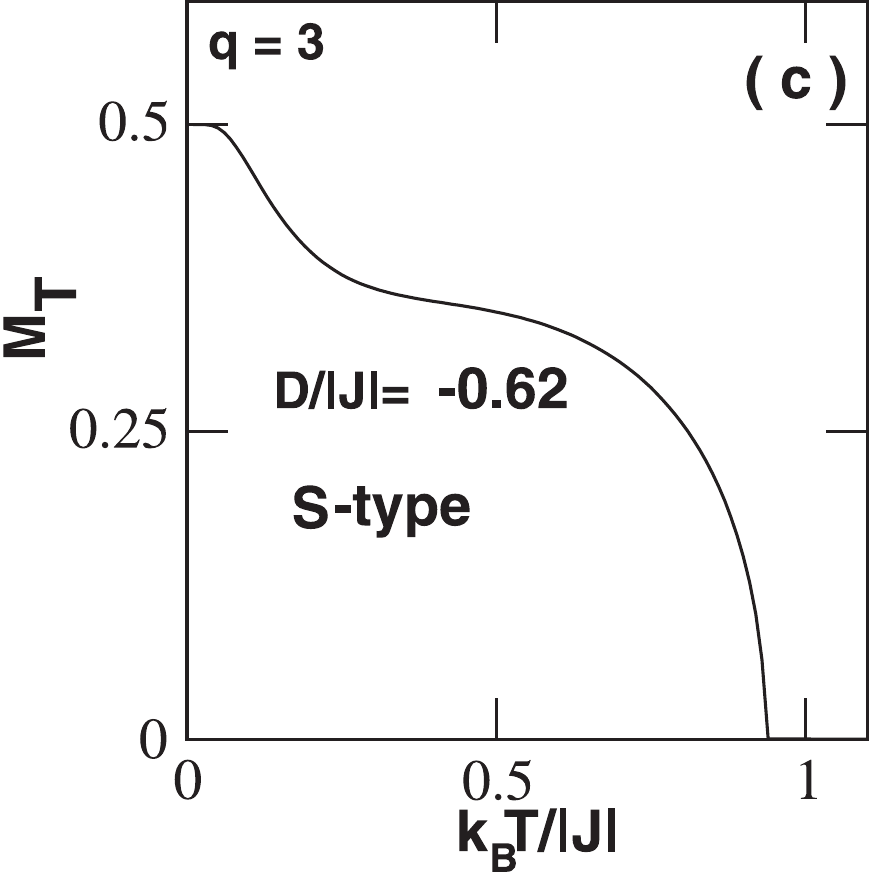}
\end{center}
\end{minipage}\\
\vspace{.2cm}
\begin{minipage}{0.33\linewidth}
\begin{center}
\includegraphics[width=\textwidth,height=5cm]{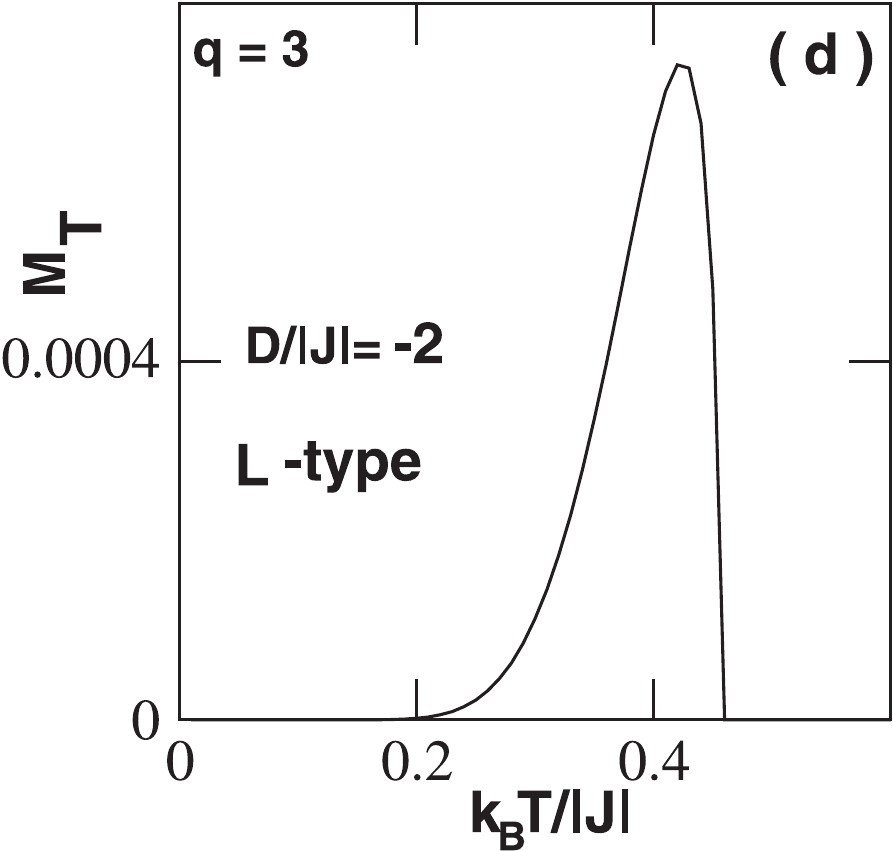}
\end{center}
\end{minipage}
\begin{minipage}{0.33\linewidth}
\begin{center}
\includegraphics[width=\textwidth,height=5cm]{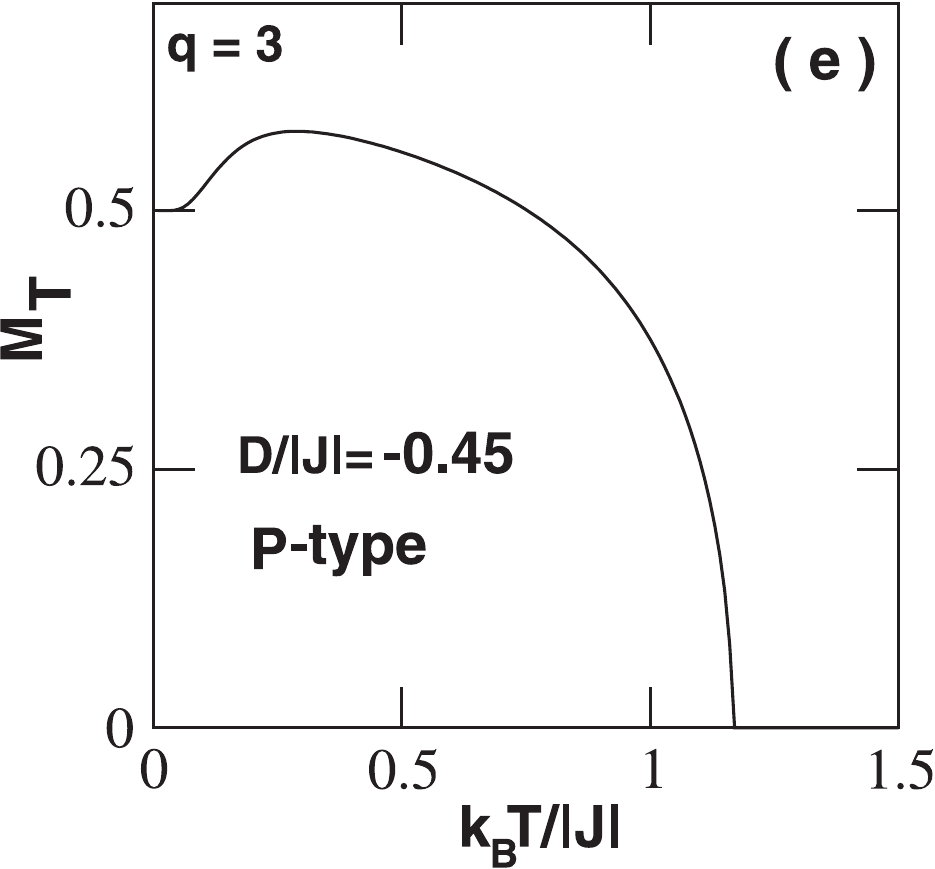}
\end{center}
\end{minipage}
\caption{
 $M_T$ as a function of the temperature for selected values of the crystal-field
 when $q=3$ as indicated in different panels. The model shows the Q-, R-, S-, L- and
P-types of compensation behaviors as classified in the extended N\'{e}el nomenclature.}
\label{fig9}
\end{figure}

In order to investigate the low-temperature magnetic properties of the model, we plotted
the sublattice magnetizations and the global magnetization at $k_{\text B} T/|J|=0.05$ for selected values
of the crystal-field as functions of the field $h$ as shown in figure~\ref{fig8}. In figure~\ref{fig8}~(a) where
$D/|J|=-1$ and $q=3$, $M_{1/2}$ shows two step-like magnetization plateaus $(M_{1/2}=-\frac{1}{2}, \frac{1}{2})$
whereas $M_{5/2}$ and $M_T$ respectively show three  and four step-like magnetization plateaus
$(M_{5/2}=\frac{1}{2}, \frac{3}{2}, \frac{5}{2})$ and $(M_T=0, \frac{1}{2}, 1, \frac{3}{2})$. Also,
from figure~\ref{fig8}~(b) where $D/|J|=0$ and $q=3$, only $M_{1/2}$ and $M_T$ present two step-like magnetization
plateaus ($M_{1/2}=-\frac{1}{2},\frac{1}{2}$) and ($M_T=1,\frac{3}{2}$). The obtained results
are consistent with the ground-state phase diagram displayed in figure~\ref{fig2}.
We also investigated the global magnetization as a function of the temperature and obtained some
compensation types of Ising model. Figure~\ref{fig9} shows temperature dependencies of the global magnetization
$M_{T}$ for selected values of the crystal-field when $q=3$. As seen in figure~\ref{fig9}, the model
exhibits five types of compensation behaviors, namely Q-, R-, S-, L- and P-type compensation behaviors as
classified in the extended N\'{e}el nomenclature \cite{ra39}.

\subsection{Finite-temperature phase diagrams}

\begin{figure}[!t]
\begin{center}
\includegraphics[angle=0,width=0.45\textwidth]{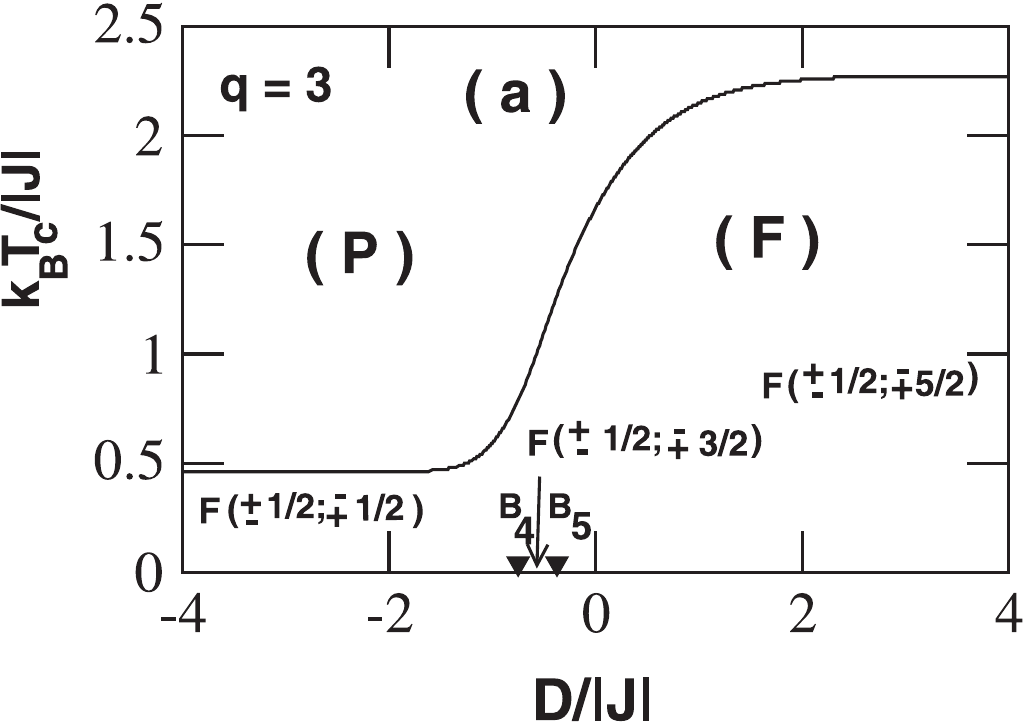} \hspace{0.8cm}
\includegraphics[angle=0,width=0.45\textwidth]{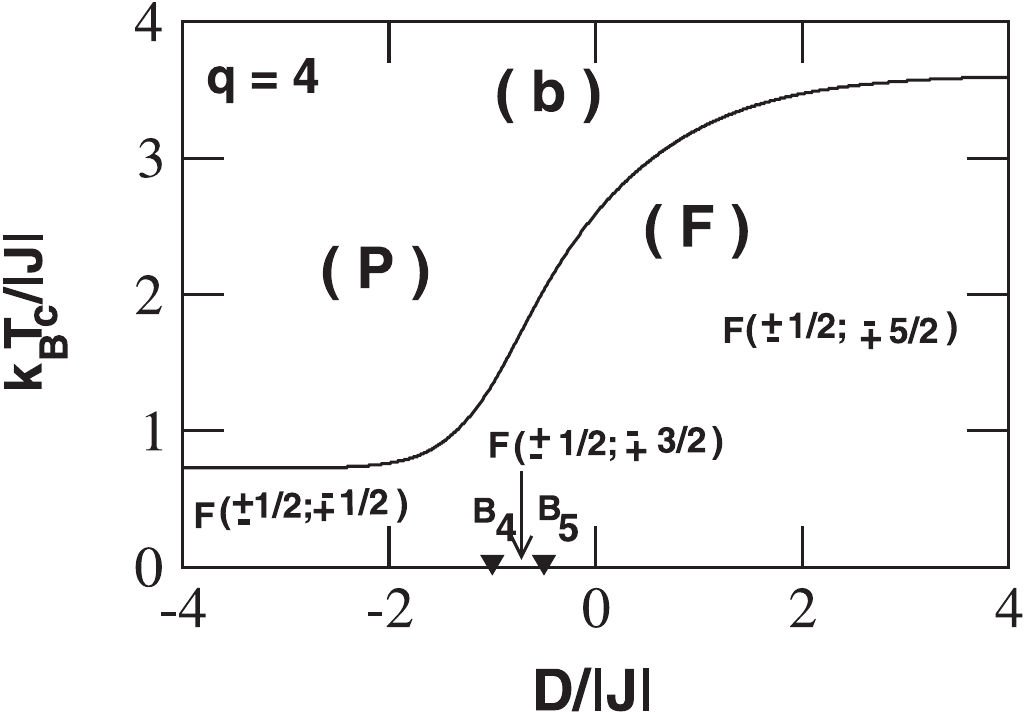}\\
\vspace{0.2cm}
\includegraphics[angle=0,width=0.45\textwidth]{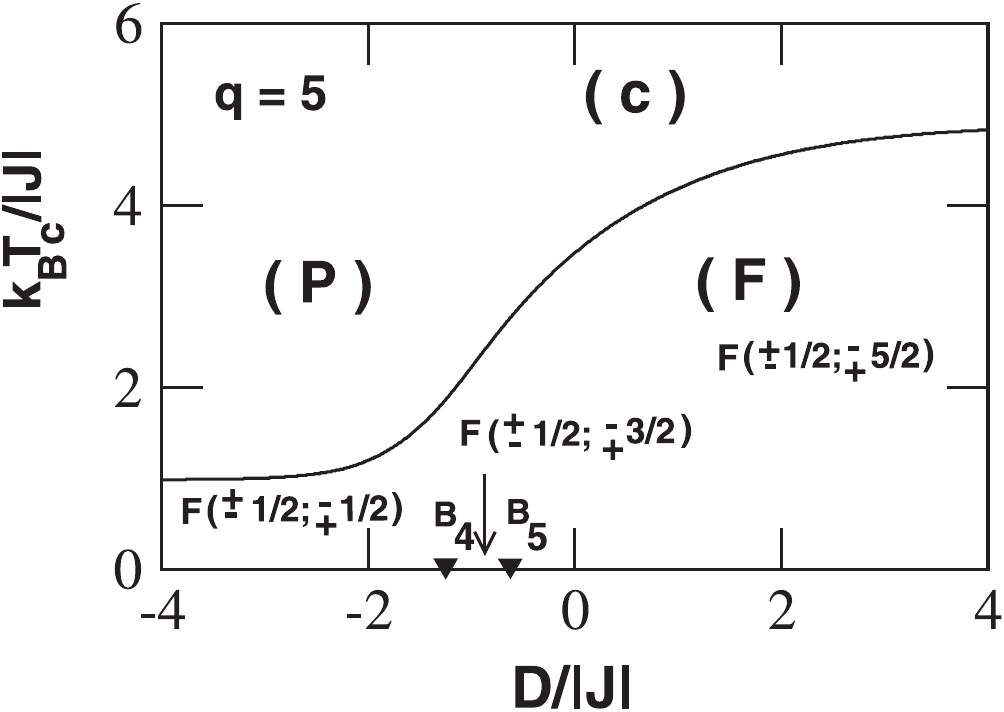} \hspace{0.8cm}
\includegraphics[angle=0,width=0.45\textwidth]{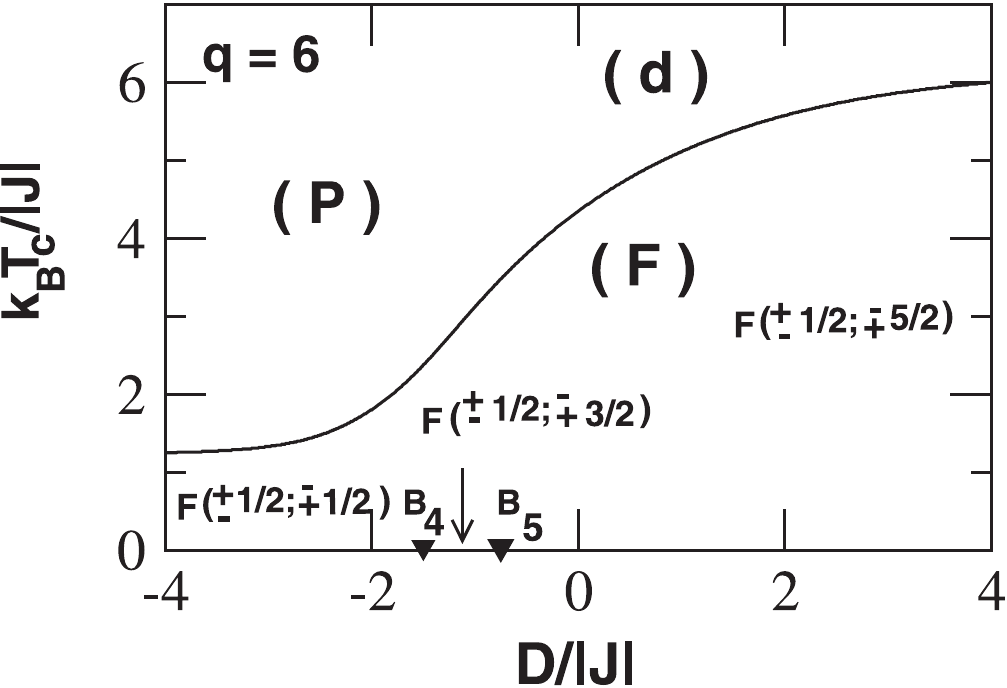}
\end{center}
\caption{
 Temperature phase diagrams of the model in the
$(D/|J|, \,k_{\textrm{B}}T/|J|)$ plane. The solid line indicates the second-order transition line.
  Panel (a): $q=3$; panel (b): $q=4$; panel (c): $q=5$ and panel (d): $q=6$.
Here, the model only presents second-order transition
 for all values of  $q$. The multicritical points $B_4$ and $B_5$ which respectively indicate
the positions of the hybrid phases $\text{ F}(\pm{\frac{1}{2}}, \mp{1})$ and $\text{ F}(\pm{\frac{1}{2}}, \mp{2})$,
respectively separate
the ferrimagnetic phases $\text{F}(\pm \frac{1}{2}, \mp \frac{1}{2})$, $\text{F}(\pm \frac{1}{2}, \mp \frac{3}{2})$
and $\text{F}(\pm \frac{1}{2}, \mp \frac{5}{2})$.}
\label{fig10}
\end{figure}

Considering all the above calculations, we can illustrate the temperature phase diagrams of
 the model. So in figure~\ref{fig10}, we have constructed the phase diagrams of the system in the $(D/|J|, \,k_{\textrm{B}}T/|J|)$
 plane in the absence of a magnetic field $h$ when $q=3, 4, 5$ and 6. In different phase
 diagrams, the solid line indicates the second-order transition line. Two filled triangles indicate
 two multicritical points $B_4$ and $B_5$ found in the ground-state phase diagram displayed in
 figure~\ref{fig2}.

 From this figure, some interesting properties of the system are singled out. Indeed, for all values of
  the coordination number $q$, from panel (a) to panel (d) where $q=3, 4, 5$ and 6, respectively, the transition lines are only of
  the second-order separating the ferrimagnetic phase (F) which is a mixture of five different ferrimagnetic
  phases from the paramagnetic phase (P) and
  become constant for $D/|J| <-q/4$. One can observe that: $(1)$ When
  $D/|J|>-q/8$, the second-order phase transition turns from ferrimagnetic phase
  $\text{F}(\pm \frac{1}{2}, \mp \frac{5}{2})$ to a
  disordered paramagnetic phase P. $(2)$ For $-q/4< D/|J|< -q/8$, the second-order
  phase transition is from the ferrimagnetic $\text{F}(\pm \frac{1}{2}, \mp \frac{3}{2})$  to the paramagnetic phase P. $(3)$ When
 $ D/|J|< -q/4$, the second-order phase transition is from the ferrimagnetic
 phase $\text{F}(\pm \frac{1}{2}, \mp \frac{1}{2})$  to the paramagnetic phase P. $(4)$~For $D/|J|=-q/8$ respectively $D/|J|=-q/4$, the
  second-order transition phase is from the hybrid phase $\text{F}(\pm{\frac{1}{2}}, \mp{2})$ respectively the hybrid
  phase $\text{ F}(\pm{\frac{1}{2}}, \mp{1})$ to the paramagnetic phase P.

  It is important to mention that figure~\ref{fig10} presents some resemblances concerning the second-order
  transition lines with figure~\ref{fig3} of \cite{ra18}. Moreover, by increasing the value of the coordination number $q$,
  the ferrimagnetic domain F becomes important.

\section{Conclusion}
\vspace{-1mm}

  In this work, the study of the mixed spin-$\frac{1}{2}$ and spin-$\frac{5}{2}$ Ising ferrimagnetic model on the Bethe lattice
  in the presence of a longitudinal magnetic field is undertaken via exact recursion equations
  method. All the thermodynamical quantities of interest are calculated as functions of recursion relations.

  The ground-state phase diagram of the model is displayed as shown in figure~\ref{fig2}. From this
  phase diagram, we have found eleven existing and stable phases and along the $D/q|J|$-axis, two particular
  hybrid phases appear at the two multicritical points $B_4$ and $B_5$.  The ground-state phase diagram is considered and used
  as a guide in obtaining different temperature phase diagrams. In the presence and in the absence of a longitudinal magnetic field $h$, we investigated thermal variations of sublattice magnetizations,
  global magnetization, the corresponding response
  functions and the internal energy as seen in figures~\ref{fig3}--\ref{fig9}. From these
  figures, the order-parameters in most cases showed a usual decay with thermal fluctuations.
  By using thermal behaviors of the considered order-parameters, and by analysing the corresponding
  response functions and the internal energy, the nature of different
  phase transitions encountered is identified. This enables us to construct and to discuss in detail
  different temperature dependence phase diagrams in the case of equal crystal-field
  interactions as shown in figure~\ref{fig10}. The model shows rich physical properties, namely
  the second-order transition and multicritical points  for all values
  of the crystal-field interactions and for all values of the coordination number $q$.

  Finally, it is useful to mention that different results found here  are compared to those reported
  in some references, and some topological similarities are shown, especially with those found in
   \cite{ra18}  where the same model is investigated
  by means of the effective-field theory with correlation. However, during our investigation we have not found
  any compensation temperature.

\ukrainianpart

\title[]
{ Вивчення   змішаної спін-$\frac{1}{2}$ і спін-$\frac{5}{2}$ моделі Ізінга на гратці Бете під дією магнітного поля}

\author[]{М.~Каріму\refaddr{label1}, Р.А.~Єссуфу\refaddr{label1,label2}, Т.Д. Оке\refaddr{label1,label2}, A.~Кпандону\refaddr{label1,label3}, Ф.~Онтінфінде\refaddr{label1,label2}}
\addresses{
\addr{label1} Інститут математики і фізичних наук (IMSP), Республіка Бенін
\addr{label2} Фізичний факультет, Університет м. Абомей-Калаві, Республіка Бенін
\addr{label3} Університет м. Наттінгу, Еколь нормаль сюпер'єр, Республіка Бенін
}

\makeukrtitle

\begin{abstract}

Досліджується змішана спін-$\frac{1}{2}$ і спін-$\frac{5}{2}$ модель на гратці Бете в присутності магнітного поля $h$, використовуючи  метод рекурсивних співвідношень. Побудована фазова діаграма основного стану, яка може бути  використана для дослідження цікавих областей температурної фазової діаграми моделі. Для того, щоб прокласифікувати  природу фазового переходу та отримати відповідні температури, детально досліджено параметри порядку, відповідні функції відгуку і внутрішню енергію.
Так, при відсутності магнітного поля, температурні фазові діаграми є представлені для випадку однакового кристалічного поля на площині
$(k_{\textrm{B}}T/|J|, D/|J|)$, коли $q=3,4,5$ і $6$. Модель демонструє тільки фазовий перехід другого роду для відповідних значень фізичних параметрів моделі.

\keywords магнітні системи, температурні зміни,  фазові діаграми, магнітне поле, перехід другого роду

\end{abstract}

\begin{thebibliography}{99}
\bibitem{ra1} Thompson C.J., Mathematical Statistical Mechanics, Princeton University Press, New Jersey, 1992.
 \bibitem{ra2} Stre\v{c}ka J., Ja\v{s}\v{c}ur M., Acta Phys. Slovaca, 2015, \textbf{65}, 235,  and references therein.
 \bibitem{ra3} White R.M., Science, 1985, \textbf{229}, 11; \bibdoi{10.1126/science.229.4708.11}.
\bibitem{ra4} Wood R., Understanding Magnetism, Tab Books Inc, Blue Ridge Summit, PA, 1988.
\bibitem{ra5} K\"{o}ster E., J. Magn. Magn. Mater., 1993, \textbf{120}, 1;  \bibdoi{10.1016/0304-8853(93)91274-B}.
\bibitem{ra6} Lueck L.B.,  Gilson R.G.,  J. Magn. Magn. Mater., 1990, \textbf{88}, 227;  \bibdoi{10.1016/S0304-8853(97)90032-9}.
\bibitem{ra7} Molecular Magnetism: New Magnetic Materials, Itoh K., Kinoshita M. (Eds.), Kodansha, Tokyo, 2000.
\bibitem{ra8} Molecular Magnets. Recent Highlights, Linert W., Verdaguer M. (Eds.), Springer-Verlag, Wien, 2003.
\bibitem{ra9} Gatteschi D.,  Adv. Mater., 1994, \textbf{6}, 635;  \bibdoi{10.1002/adma.19940060903}.
\bibitem{ra10} Miller J.S.,  Epstein A.J.,  Chem. Eng. News, 1995, \textbf{73}, 30;  \bibdoi{10.1021/cen-v073n040.p030}.
\bibitem{ra11} Kahn O., Molecular Magnetism, VCH, New York, 1993.
\bibitem{ra12} Bob\'{a}k A., Physica A, 1998, \textbf{258}, 140;  \bibdoi{10.1016/S0378-4371(98)00233-7}.
\bibitem{ra13} Benyoussef A., El Kenz A., Kaneyoshi T., J. Magn. Magn. Mater., 1994, \textbf{131}, 179;  \bibdoi{10.1016/0304-8853(94)90026-4}.
\bibitem{ra14} Bob\'{a}k A., Jur\u{c}i\v{s}in M., Physica A, 1997, \textbf{240}, 647;  \bibdoi{10.1016/S0378-4371(97)00044-7}.
\bibitem{ra15} De Oliveira D.C., Silva A.A.P., de Albuquerque D.F., de Arruda A.S., Physica A, 2007, \textbf{386}, 205; \\ \bibdoi{10.1016/j.physa.2007.07.073}.
\bibitem{ra16} Kaneyoshi T., Physica A, 1994, \textbf{205}, 677;  \bibdoi{10.1016/0378-4371(94)90229-1}.
 \bibitem{ra17} Guo K.T., Xiang S.H., Xu H.Y., Li X.L., Quantum Inf. Process., 2014, \textbf{13}, 1511;  \bibdoi{10.1007/s11128-014-0745-7}.
 \bibitem{ra18} Deviren B., Keskin M., Canko M.O., Physica A, 2009, \textbf{388}, 1835;  \bibdoi{10.1016/j.physa.2009.01.032}.
 \bibitem{ra19} Da Cruz Filho J.S.,  Godoy M., de Arruda A.S., Physica A, 2013, \textbf{392}, 6247;  \bibdoi{10.1016/j.physa.2013.08.007}.
 \bibitem{ra20} Miao H.,  Wei G., Geng J., J. Magn. Magn. Mater., 2009, \textbf{321}, 4139;  \bibdoi{10.1016/j.jmmm.2009.08.018}.
 \bibitem{ra21} Mohamad H.K., Domashevskaya E.P., Klinskikh A.F., Physica A, 2009, \textbf{388}, 4713;  \bibdoi{10.1016/j.physa.2009.08.014}.
 \bibitem{ra22} Mohamad H.K., Int. J. Adv. Res., 2014, \textbf{2}, No. 9, 442.
\bibitem{ra23} Kaneyoshi T., Chen J.C., J. Magn. Magn. Mater., 1991, \textbf{98}, 201;  \bibdoi{10.1016/0304-8853(91)90444-F}.
\bibitem{ra24}  Quadros S.G.A., Salinas S.R., Physica A, 1994, \textbf{206}, 479;  \bibdoi{10.1016/0378-4371(94)90319-0}.
\bibitem{ra25}   El Bouziani M., Gaye A.,  Jellal A., Physica A, 2013, \textbf{392}, 689;  \bibdoi{10.1016/j.physa.2012.10.007}.
\bibitem{ra26}  Buendia G.M.,  Liendo J.A., J. Phys.: Condens. Matter, 1997, \textbf{9}, 5439;  \bibdoi{10.1088/0953-8984/9/25/011}.
\bibitem{ra27}  Godoy M., Figueiredo W., Phys. Rev. E, 2002,  \textbf{66}, 036131;  \bibdoi{10.1103/PhysRevE.66.036131}.
\bibitem{ra28}  Cambui D.S.,  Arruda A.S.,  Godoy M., Int. J. Mod. Phys. C, 2012, \textbf{23}, 1240015;  \bibdoi{10.1142/S0129183112400153}.
\bibitem{ra29}  Feraoun A.,  Zaim A.,  Kerouad M.,  Physica B, 2014, \textbf{445}, 74;  \bibdoi{10.1016/j.physb.2014.03.071}.
\bibitem{ra30}  \u{Z}ukovi\u{c} M.,  Bob\'{a}k A., J. Magn. Magn. Mater., 2010, \textbf{322}, 2868;  \bibdoi{10.1016/j.jmmm.2010.04.043}.
\bibitem{ra31}  Yessoufou R.A.,  Bekhechi S.,  Hontinfinde F., Eur. Phys. J. B, 2011, \textbf{81}, 137;  \bibdoi{10.1140/epjb/e2011-10825-7}.
\bibitem{ra32}  Kple J.,  Yessoufou R.A.,  Hontinfinde F.,  Afr. Rev. Phys., 2012, \textbf{7}, 319.
\bibitem{ra33}  Yigit A., Albayrak E., Chinese Phys. B,  2012, \textbf{21}, 020511;  \bibdoi{10.1088/1674-1056/21/2/020511}.
\bibitem{ra34}  Ekiz C., Phys. Lett. A, 2007, \textbf{367}, 483;  \bibdoi{10.1016/j.physleta.2007.03.038}.
\bibitem{ra35}  Albayrak E.,  Yigit A.,  Phys. Lett. A,  2006, \textbf{353}, 121;  \bibdoi{10.1016/j.physleta.2005.12.077}.
\bibitem{ra36}  Karimou M.,  Yessoufou R.,  Hontinfinde F.,  Int. J. Mod. Phys. B, 2015, \textbf{29}, 1550194;  \bibdoi{10.1142/S0217979215501945}.
\bibitem{ra37}  Ekiz C., J. Magn. Magn. Mater., 2006, \textbf{307}, 139;  \bibdoi{10.1016/j.jmmm.2006.03.059}.
\bibitem{ra38}  Ekiz C., Commun. Theor. Phys., 2009, \textbf{52}, 539;  \bibdoi{10.1088/0253-6102/52/3/30}.
\bibitem{ra39}  Ekiz C.,  Stre\v{c}ka J., Ja\v{s}\u{c}ur M., Cent. Eur. J. Phys., 2009, \textbf{7}, 509;  \bibdoi{10.2478/s11534-009-0043-7}, and references therein.
\end{thebibliography}
\end{document}